\documentclass[runningheads]{llncs}
\usepackage{graphicx}
\usepackage{siunitx}
\usepackage{color}
\usepackage{amssymb}
\usepackage{makecell}
\usepackage{multirow}
\usepackage{lipsum}
\usepackage{caption}
\usepackage{subcaption}
\usepackage{adjustbox}
\usepackage{comment}
\usepackage{float}
\usepackage{booktabs}
\usepackage{rotating}
\usepackage{setspace}
\usepackage{starfont}

\newcommand{\ra}[1]{\renewcommand{\arraystretch}{#1}}

\usepackage{pifont}
\definecolor{MyBlue}{rgb}{0,0.08,0.5}
\definecolor{MyRed}{rgb}{0.7,0.02,0.02}
\definecolor{MyOrange}{rgb}{1,0.5,0}
\definecolor{MyPurple}{rgb}{0.6,0.25,0.8}
\definecolor{MyGreen}{rgb}{0.1,0.8,0.1}

\newcommand{\etal}{\textit{et al}.}

\usepackage[pagebackref=true,breaklinks=true,letterpaper=true,colorlinks=false,bookmarks=false]{hyperref}
\hypersetup{
     colorlinks = true,
     linkcolor = red,
     anchorcolor = black,
     citecolor = black,
     filecolor = black,
     urlcolor = blue
     }

\begin{document}
\title{Stable deep neural network architectures for mitochondria segmentation on electron microscopy volumes}
\titlerunning{Stable Deep Neural Network architectures}

\newcommand*{\affaddr}[1]{#1} 
\newcommand*{\affmark}[1][*]{\textsuperscript{#1}}

\author{Daniel Franco-Barranco\affmark[1,2] \and Arrate Muñoz-Barrutia\affmark[3,4] \and Ignacio Arganda-Carreras\affmark[1,2,5]  }

\authorrunning{Franco-Barranco {\em et al.}}
\institute{
\affaddr{\affmark[1] University of the Basque Country (UPV/EHU)} \\
\affaddr{\affmark[2] Donostia International Physics Center (DIPC)} \\
\affaddr{\affmark[3] Universidad Carlos III de Madrid} \\
\affaddr{\affmark[4] Instituto de Investigación Sanitaria Gregorio Marañón} \\
\affaddr{\affmark[5] Ikerbasque, Basque Foundation for Science} \\
\email{daniel.franco@dipc.org}
}

\maketitle   
Electron microscopy (EM) allows the identification of intracellular organelles such as mitochondria, providing insights for clinical and scientific studies. In recent years, a number of novel deep learning architectures have been published reporting superior performance, or even human-level accuracy, compared to previous approaches on public mitochondria segmentation datasets. Unfortunately, many of these publications do not make neither the code nor the full training details public to support the results obtained, leading to reproducibility issues and dubious model comparisons. For that reason, and following a recent code of best practices for reporting experimental results, we present an extensive study of the state-of-the-art deep learning architectures for the segmentation of mitochondria on EM volumes, and evaluate the impact in performance of different variations of 2D and 3D U-Net-like models for this task. To better understand the contribution of each component, a common set of pre- and post-processing operations has been implemented and tested with each approach. Moreover, an exhaustive sweep of hyperparameters values for all architectures have been performed and each configuration has been run multiple times to report the mean and standard deviation values of the evaluation metrics. Using this methodology, we found very stable architectures and hyperparameter configurations that consistently obtain state-of-the-art results in the well-known EPFL Hippocampus mitochondria segmentation dataset. Furthermore, we have benchmarked our proposed models on two other available datasets, Lucchi++ and Kasthuri++, where they outperform all previous works. The code derived from this research and its documentation are publicly available at \url{https://github.com/danifranco/EM_Image_Segmentation}.
\section{Introduction}

Recent imaging methods in electron microscopy (EM) allow scientists to identify subcellular organelles such as vesicles or mitochondria with nano-scale precision. In particular, mitochondria play an important role in some crucial functions in the cell, such as energy production, signaling, differentiation, cell growth and death~\cite{tait2012mitochondria}. For that reason, the automated and accurate segmentation of mitochondria is specially relevant for basic research in neuroscience, but in clinical studies as well, since their number and morphology are related to severe diseases such as cancer~\cite{de2010mitochondrial,fulda2010targeting,wallace2012mitochondria}, Parkinson disease~\cite{poole2008pink1} or Alzheimer disease~\cite{de2010mitochondrial}.

In the past decade, advances in machine learning and computer vision, especially those based on deep learning, have helped scientists to automatically quantify the size and morphology of cells and organelles in microscopy images~\cite{moen2019deep,meijering2020bird}. However, with an increasing number of deep learning-based bioimage segmentation publications every year, there is a lack of enough benchmarks for different image modalities and segmentation problems to compare state-of-the-art methods under the same conditions. Moreover, deep learning methods are usually too data-specialized, making it difficult to identify those approaches that perform well on datasets different from those they have been tested on~\cite{isensee2021nnu}. On top of that, many of such approaches are published without their supporting code and image data, leading to major reproducibility and reliability problems. Such issues have not gone unnoticed and are increasingly in focus. They have become the main target even for recently proposed challenges\footnote{\url{https://paperswithcode.com/rc2020}} where the machine learning community aims at reproducing the computational experiments and verifying the empirical results already published at top venues.

As pointed out by recent works~\cite{bello2021revisiting,isensee2021nnu}, while many publications insist on presenting architectural novelties, the overall performance of a network depends substantially on its corresponding pre-processing, training, inference and post-processing strategies. Even though such choices play a critical role in the final results, very often they tend to be omitted in the method descriptions and their comparisons with competing approaches.

Another issue inherent to the use of deep learning architectures (and frequently not discussed in publications) is the sometimes not negligible variability of the results produced by different executions of the same exact architecture and training configuration. Despite programmatically setting all initial random seeds, the non deterministic nature of the graphical processing units (GPUs) introduces variations from execution to execution, resulting on slightly different performances. This variability is usually not taken into account when presenting results, although it could be crucial to select models, training and inference strategies that repeatedly lead to stable results.

In the particular task of mitochondria segmentation, the \textit{de facto} benchmark dataset adopted by the community is the EPFL Hippocampus dataset~\cite{lucchi2011supervoxel} (hereafter referred to as Lucchi dataset). Published in 2011, it contains two image volumes (training and test) of the same size, and their respective semantic segmentation labels are both public. As the reference in the field for a decade, many methods have been published proposing solutions for this dataset. Unfortunately, most of them suffer from the aforementioned problems, forcing other scientists to code their own versions of the published algorithms, many times knowing too few details about their original implementations, training and inference methodologies.



To address this deficiency in the field, we first re-implemented the top-performing deep learning architectures for the Lucchi dataset following the descriptions of their original publications. None of them led directly to their claimed results. After our own modifications, an extensive hyperparameter search and multiple runs of the same configuration, some of these methods occasionally achieved such results. Second, we compared the performance of state-of-the-art biomedical semantic segmentation architectures in the same dataset, evaluated under the same training and inference framework. In particular, we focused on the stability of the resulting metric values after several executions of the same configuration and scrutinized the impact of different popular post-processing and output reconstruction methods. Finally, based on our findings, we propose light encoder-decoder architectures that consistently lead to robust state-of-the-art results in the Lucchi dataset as well as in other public mitochondria segmentation datasets.

\ \

In brief, our main contributions are as follows:
\begin{enumerate}
    \item We performed a thorough study on the reproducibility and stability of the top-performing deep learning segmentation methods published for the Lucchi dataset, exposing major issues to achieve their claimed results in a consistent manner.
    \item We made a comprehensive comparison of the performance of the most popular deep learning architectures for biomedical segmentation using the Lucchi dataset, and show their stability under the same training and post-processing conditions. 
    \item We propose different variations of light-weight encoder-decoder architectures, together with a training/inference workflow, that lead to stable and robust results across mitochondria segmentation datasets. 
\end{enumerate}

The rest of this paper is structured as follows. In Section~\ref{related}, we review the state of the art in biomedical semantic segmentation, with a special focus on mitochondria segmentation. In Section~\ref{methodology}, we introduce our proposed architectures and the different post-processing and test-time evaluation methods. In Section~\ref{experiments}, we introduce the datasets and evaluation metrics employed. We also show the results that support our findings together with an ablation study that unveils the contribution of every component of our proposed solution. Finally, in Section \ref{conclusions}, the conclusions of this work are presented.

\section{Related work}
\label{related}

In the last decade, deep learning approaches have become dominant in the field of computer vision and its most common target applications~\cite{garcia2018survey,minaee2021image} including semantic segmentation for biomedical image analysis~\cite{haque2020deep,litjens2017survey}. In particular, the semantic segmentation problem aims at linking each pixel in an image to a class label, producing an output of the same size as the input image. The first steps towards resolving this problem using deep learning were taken by means of fully convolution networks (FCNs)~\cite{long2015fully}. More specifically, fully connected layers were replaced by convolutional layers in some classic networks such as AlexNet~\cite{krizhevsky2012imagenet}, VGG~\cite{simonyan2014very} or GoogLeNet~\cite{szegedy2015going} and information from intermediate layers was fused to upsample the feature maps encoded by the network, finally producing a pixel-wise classification. This idea of \textit{encoding} the image through a convolutional neural network (CNN), outputing a vector feature map (also called \textit{bottleneck}), and recovering its original spatial shape in a \textit{decoding} path was further extended in subsequent works~\cite{noh2015learning,ronneberger2015u,milletari2016v,jegou2017one,badrinarayanan2017segnet,chaurasia2017linknet}.

A major breakthrough was the U-Net~\cite{ronneberger2015u}, that extended the encoding and decoding idea by making an upsampling path with up-convolutions after the bottleneck to recover the original image size. In addition, the authors proposed skip connections between the contracting and the expanding path, allowing the upsampling path to recover fine-grained details. The U-Net is the baseline of numerous approaches due to its success on multiple biomedical applications. For instance, U-Net-based methods have been proposed to segment, among others, polyps in colonoscopy videos~\cite{zhou2018unet++}, liver in abdominal computed tomography (CT) scans~\cite{zhou2018unet++}, pancreas in abdominal CT
scans~\cite{SCHLEMPER2019197}, brain magnetic resonance imaging scans~\cite{roy2018concurrent}, neurites~\cite{arganda2015crowdsourcing,gu2019net} and synapses in brain EM images~\cite{buhmann2018synaptic}, dermoscopy images~\cite{ibtehaz2020multiresunet} or retina blood vessels in diabetic retinopathy images~\cite{zhuang2018laddernet,jin2019dunet}.

In the specific case of mitochondria segmentation, early works attempting to segment the Lucchi dataset~\cite{lucchi2011supervoxel} leveraged traditional image processing and machine learning techniques~\cite{lucchi2012structured,lucchi2013learning,lucchi2014exploiting,lucchi2014learning}. In their last two works, Lucchi \etal~proposed alternative methodologies to segment mitochondria on their own dataset explicitly modeling mitochondria membranes~\cite{lucchi2014exploiting,lucchi2014learning}. From that last work, Casser \etal~\cite{casser2020fast} inferred a Jaccard index or intersection over the union (IoU) lower bound value of $0.895$ in the test set. The IoU is a common way of measuring the overlapping area between the ground truth and the produced segmentation with values that range from $0$ to $1$, where $0$ means no overlap at all and $1$ represents a perfect match (see Section~\ref{sec:eval_metrics}).  

More modern approaches made use of deep learning architectures to segment the Lucchi dataset. For instance, Oztel \etal~\cite{oztel2017mitochondria} trained a CNN with four convolutional layers to classify $32\times32$ pixel patches extracted from the training data into two classes: mitochondria and background. After that, they fed the network with the full test images to simulate a sliding window process and applied three consecutive post-processing methods to improve the segmentation results: 1) spurious detection to remove small false blobs, 2) marker-controlled watershed transform~\cite{meyer1994topographic} for border refinement, and 3) a median filtering to smooth labels along the z-axis. This way, they reported an IoU value of $0.907$ in the test set, which is the highest reported value to date. Liu \etal~\cite{liu2018automatic} used instead a modified Mask Region-based CNN (Mask RCNN~\cite{he2017mask}) to detect and segment mitochondria. As post-processing methods they performed: 1) a morphological opening operation to eliminate small regions and smooth large ones, 2) a multi-layer fusion operation to exploit 3D mitochondria information, and 3) a size-based filtering to remove tiny segmented objects that have an IoU score below a given threshold. As a result, they reported an IoU value of $0.849$ in the test set. Cheng \etal~\cite{cheng2017volume} applied both a 2D and a 3D version of an asymmetric U-Net-like network. They introduced the \textit{stochastic downsampling} method, that produces augmentation at feature-level in an operation they named \textit{feature level augmentation}. More specifically, on that downsampling layer, they subdivided the image in fixed square regions and picked random rows and columns inside them to select the pixels/voxels that will constitute the downsampled output. Moreover, they implemented factorized convolutions~\cite{szegedy2016rethinking} instead of classical ones to drastically reduce the number of network parameters. As best result, they reported an IoU value of $0.889$ in the test set using their 3D network. Xiao \etal ~in~\cite{xiao2018automatic} employed a variant of a 3D U-Net model with residual blocks. Additionally, in the decoder of the network they included two auxiliary outputs to address the vanishing gradient issue. Their final output is the result of the ensemble prediction of the 16 possible 3D variations (using flips and axis rotations) per each 3D subvolume. They reported an IoU value of $0.900$ in the test set. Finally, in a more recent work, Casser \etal~\cite{casser2020fast} presented a light version of a 2D U-Net aiming to achieve real-time segmentation. They reduced the number of network parameters to $2$M and reported an IoU value of $0.890$ applying median \textit{Z-filtering} as post-processing method as well.

\section{Methodology}
\label{methodology}

Although an increasing number of methods are published every year proposing architectural modifications of a basic U-Net to perform biomedical segmentation, it is usually unclear if their claimed superiority is only due to an incomplete optimization of the basic network for the task at hand~\cite{isensee2021nnu,bello2021revisiting}. We hypothesize that, on top of answering that question, a full optimization can also lead to lightweight models that constantly produce stable and robust results across different datasets. To prove it, we explored basic U-Net configurations together with popular architectural tweaks such as residual connections~\cite{he2016deep} or attention gates~\cite{SCHLEMPER2019197}. Additionally, to disentangle the impact of each training choice, all configurations are run several times and their results are shown in the context of different post-processing and output reconstruction methods.

\subsection{Proposed networks}
\label{sec:proposed_networks}

\begin{figure}[t]
 \includegraphics[width=1\textwidth]{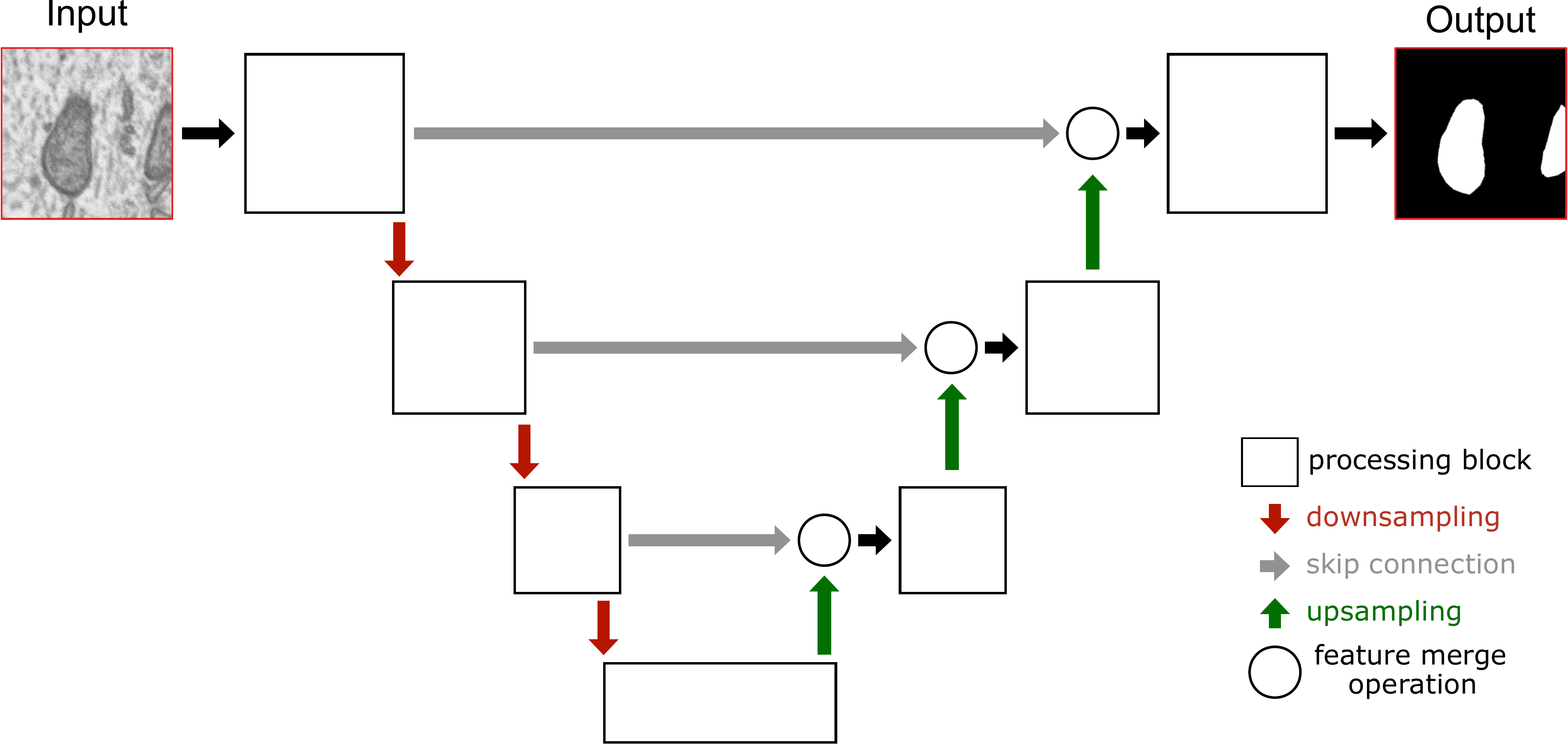}
   \caption{Graphical representation of the proposed network architectures. Depending on the model of choice, the processing blocks can be either simply convolutional or residual blocks, while the feature merge operations may imply a single concatenation or an additional attention gate.}
    \label{fig:proposed_networks}
\end{figure}

\begin{figure}[]
\centering
\begin{subfigure}{0.38\textwidth}
  \includegraphics[width=\linewidth]{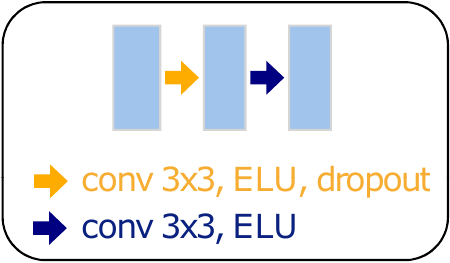}
  \caption{ }
  \label{fig:conv_block}
\end{subfigure}\hfil 
\begin{subfigure}{0.45\textwidth}
  \includegraphics[width=\linewidth]{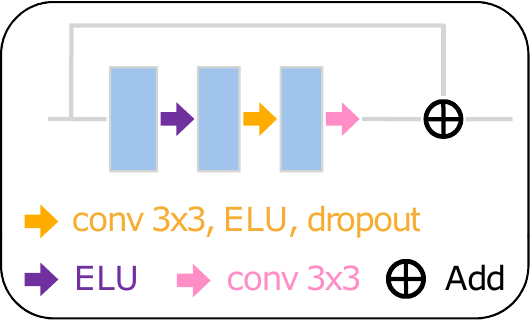}
  \caption{ }
  \label{fig:res_block}
\end{subfigure}
\caption{Types of processing blocks: (a) convolutional block used in U-Net and Attention U-Net architectures, and (b) residual block used in Residual U-Net.}
\end{figure}

Building upon the state of the art, we have explored different lightweight U-Net-like architectures in 2D and 3D. The general scheme for all architectures is represented in Figure~\ref{fig:proposed_networks}, where our basic and Attention U-Net models use convolutional blocks as processing blocks (two $3\times3$ convolutional layers, see Figure~\ref{fig:conv_block}) and our Residual U-Net is formed by full pre-activation~\cite{he2016identity} residual blocks (two $3\times3$ convolutional layers with a shortcut as shown in Figure~\ref{fig:res_block}). Both basic U-Net and Residual U-Net use concatenation as feature merge operation while our Attention U-Net introduces there an attention gate~\cite{SCHLEMPER2019197}. 

\begin{figure}[t]
\centering
     \includegraphics[width=\textwidth]{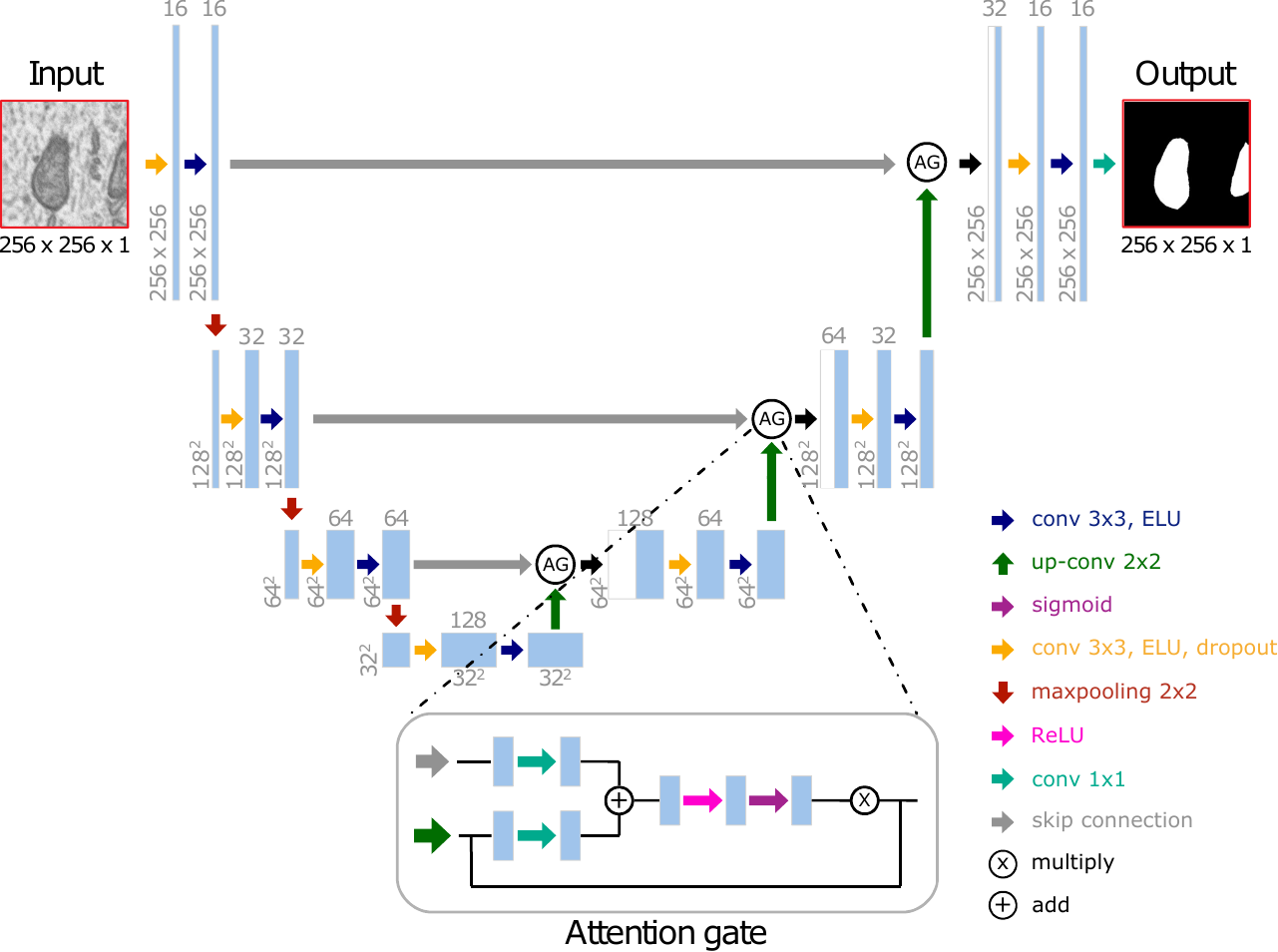} 
\caption{Example of 2D Attention U-Net architecture with 3 downsampling levels and detailed description of the attention gates used in the skip connections.}
\label{fig:attention_unet}
\end{figure}

Next, we describe the best configuration found for each architecture, based in the thorough hyperparameter exploration described in the appendixes. Namely, the best performing architectures are as follows:

\begin{itemize}
    \item \textbf{Basic U-Net}. In 2D, it is a 4-level U-Net with $16$ filters in the initial level (reducing the amount of trainable parameters from over $31$M in the original architecture~\cite{ronneberger2015u} to less than $2$M) that get double on each level, dropout in each block (from $0.1$ up to $0.3$ in the bottleneck and reversely, from $0.3$ to $0.1$ in the upsampling layers), ELU activation functions and transposed convolutions to perform the upsampling in the decoder.
    
    In 3D, the architecture is very similar, but using depth $3$, with $28$, $36$, $48$ and $64$ (in the bottleneck) 3D filters on each layer, what reduces the amount of trainable parameters to less than $0.8$M. 
    
    \item \textbf{Residual U-Net}. In 2D, this network is identical to our best basic U-Net architecture but swapping each convolutional block by a residual block~\cite{he2016deep}.
    
    For the 3D residual approach, we achieved our best results going one level deeper than the non-residual 3D network. The number of feature maps used in this case are $28$, $36$, $48$, $64$ and $80$ (bottleneck).
    
    \item \textbf{Attention U-Net}. These networks follow our 2D/3D basic U-Net architectures but incorporate attention gates~\cite{SCHLEMPER2019197} in the features passed by the skip connections. Such attention mechanism emphasizes salient feature maps that are in charge of the class decision and suppress irrelevant ones. This technique endows the network with the ability to focus on relevant regions of the image useful for a specific task. A full 3-level Attention U-Net is depicted in Figure~\ref{fig:attention_unet}. 
\end{itemize}

\subsection{Post-processing}

\begin{figure}[t]
     \centering
     \includegraphics[width=0.8\textwidth]{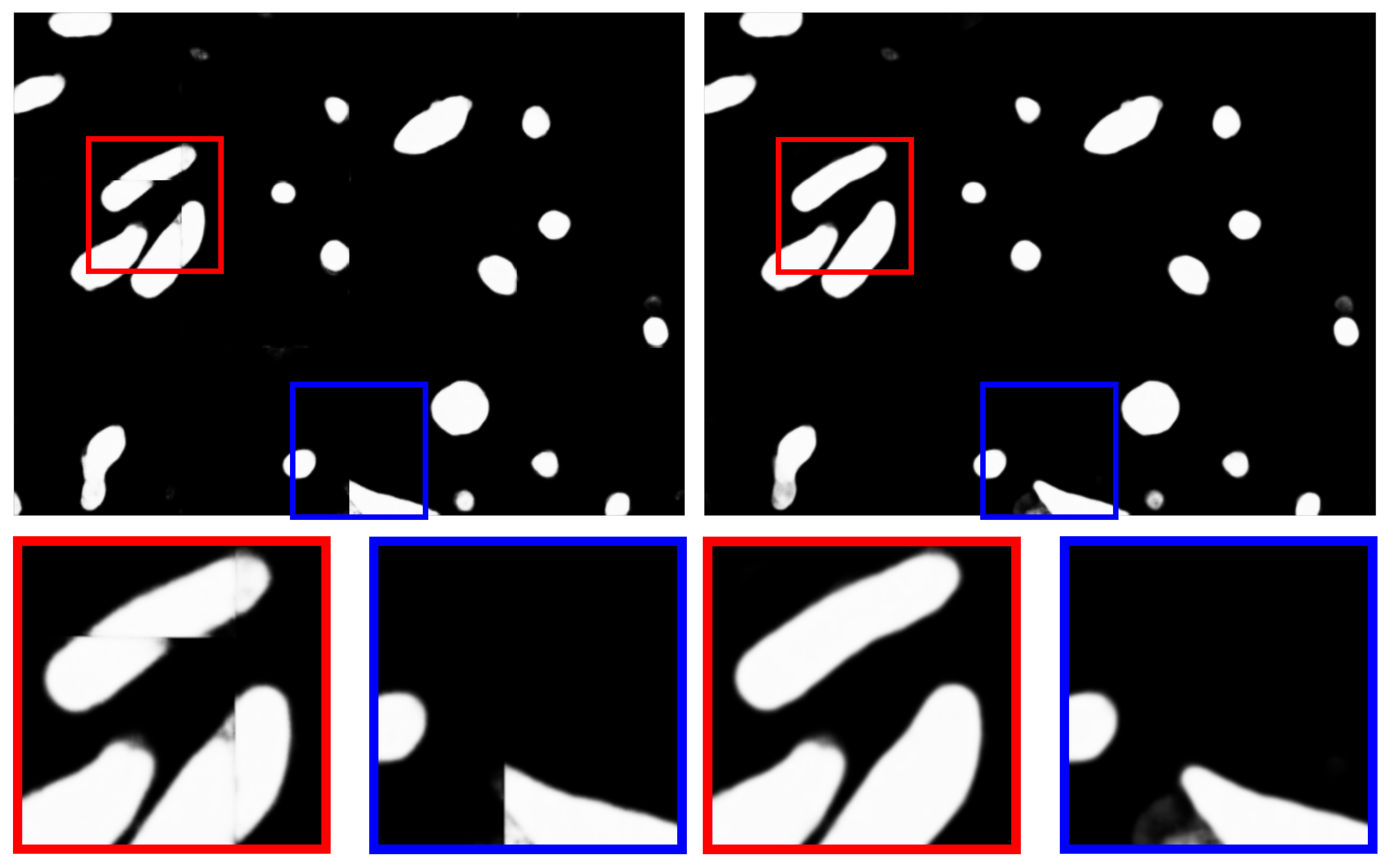}
       \caption{Border effect in output image reconstruction. From left to right: output image reconstructed from patches with visible jagged predictions; and output image reconstructed using both the blending and ensemble techniques. Blue and red boxes show zoomed areas on both images.}
    \label{fig:jag_pred}
\end{figure}

As the network outputs are pixel-wise predictions, it is common practice to apply basic post-processing methods as an easy way to improve the results. In particular, we experimented with three different techniques and studied their impact in the final segmentation result:

\begin{itemize}
    \item \textbf{Ensemble estimation.} Following a popular test-time data augmentation strategy to boost model performance, inference is applied on the multiples of $90^\circ$ rotations and flipped versions of each image. Consequently, eight versions are created in 2D and $16$ versions in 3D. Finally, the individual transformations are undone and the results are averaged into a final prediction for ensemble effect.
    
    \item \label{sec:smooth} \textbf{Blending overlapped patches.} When networks work on image patches instead of full images, the final prediction is reconstructed as a mosaic of the predictions of the patches. A recurrent problem of this strategy, as shown in Figure \ref{fig:jag_pred}, is that inference may produce jagged predictions on the borders of the predicted images. To solve this, a common approach consist on creating overlapping patches and smoothly blending the resulting predictions using a second order spline window function\footnote{\url{https://github.com/Vooban/Smoothly-Blend-Image-Patches}}. Due to its computational cost, we only experimented with this technique in 2D.
    
    \item \textbf{Median Z-filtering.} A simple median filter along the Z-axis~\cite{casser2020fast,oztel2017mitochondria} can be used to correct label predictions in consecutive image slices.
\end{itemize}

\subsection{Output reconstruction}
\label{sec:eval_uncertainties}

During the training of deep networks, the input images are commonly divided into patches due to GPU memory limitations. Therefore, those patches need to be merged back together to form the final output at full-image size. In some publications, the authors specify clearly the way they infer and merge their predictions~\cite{xiao2018automatic}, while in others this process is not described at all~\cite{cheng2017volume,oztel2017mitochondria,casser2020fast}, hindering a direct comparison between methods' performance. As it shown in Section \ref{experiments}, the evaluation scores obtained vary substantially depending on the reconstruction strategy. For that reason, and following the code of good practices to show deep learning-based results proposed by Dodge \etal~\cite{dodge2019show}, all results presented in this paper state the reconstruction strategy used. Namely, the implemented options are as follows:

\begin{enumerate}
    \item \textbf{Per patch:} The metric value corresponds to the average value over all patches. 
    \item \textbf{Per image (no overlap):} The patches are merged into a mosaic without overlap and the metric value is the average over the fully reconstructed images. Notice this option is only possible if the image size is a multiple of the patch size.
    \item \textbf{Per image (with $50\%$ overlap):} The patches are merged together using $50\%$ of overlap and the metric value is the average over all reconstructed images. 
    \item \textbf{Full image:} Inference is applied on the full-sized images and the metric value is the average over all images. Notice this strategy is not always feasible, since it depends on the input image size and the available GPU memory.
\end{enumerate}

\section{Experiments}

In order to test our hypothesis and focusing on model reproducibility and stability, we conducted a thorough study on the top-performing segmentation methods recently published in the Lucchi dataset. Additionally, we introduce our own solutions, compare them with state-of-the-art approaches in biomedical semantic segmentation, and test them in other public datasets. In all our experiments, we present average scores obtained running the same configuration $10$ times (hereafter referred as a \textit{run}) together with the corresponding standard deviation.

\label{experiments}

\subsection{Datasets}

\begin{figure}[t]
     \centering
     \includegraphics[width=0.8\textwidth]{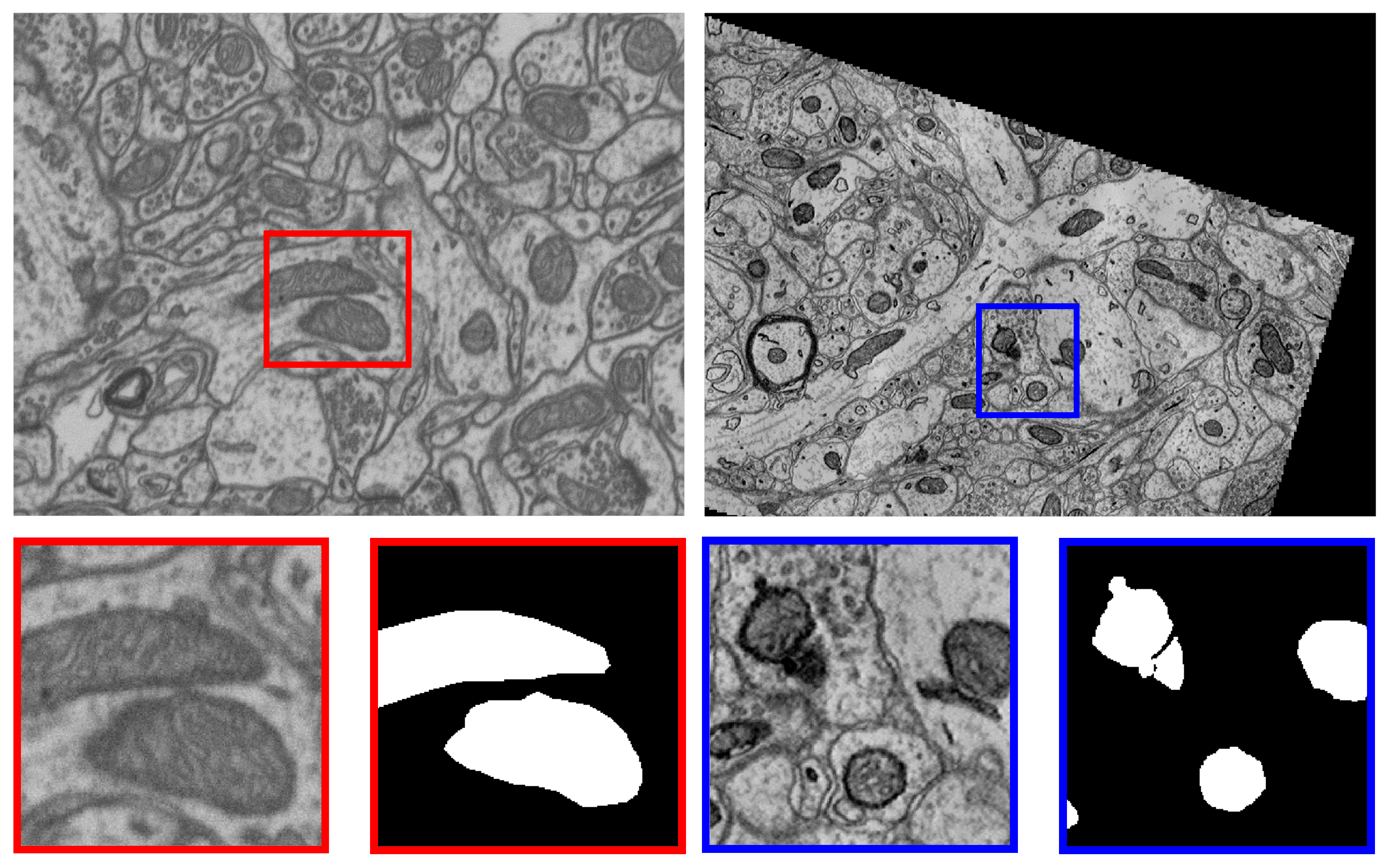}
       \caption{Sample images from public mitochondria datasets. From left to right: Lucchi and Kasthuri++ data sample with their corresponding binary mask. Blue and red boxes show zoomed areas on both images.}
    \label{fig:datasets}
\end{figure}

All the experiments performed in this work are based on the following publicly available datasets:

\textbf{EPFL Hippocampus or Lucchi dataset.} Introduced by Lucchi \etal~\cite{lucchi2011supervoxel}, this dataset has since become the \textit{de facto} standard for mitochondria segmentation in EM. The original volume represents a $5\times5\times5$ $\mu m$ section of the CA1 hippocampus region of a mouse brain, with an isotropic resolution of $5\times5\times5$ nm per voxel. The volume of $2048\times1536\times1065\:\mbox{voxels}$ was acquired using focused ion beam scanning electron microscopy (FIB-SEM). The mitochondria of two subvolumes formed by $165$ images of $1024\times768$ pixels were manually labeled by experts, and are commonly used as training and test data. An image sample is presented in Figure \ref{fig:datasets} (red).

\textbf{Lucchi++ dataset.} Presented by Casser \etal~\cite{casser2020fast}, this is a version of the Lucchi dataset after two neuroscientists and a senior biologist re-labeled mitochondria by fixing misclassifications and boundary inconsistencies.

\textbf{Kasthuri++ dataset.} Also presented by Casser \etal~\cite{casser2020fast}, this is a re-labeling of the mouse cortex dataset by Kasthuri \etal~\cite{kasthuri2015saturated}. The volume corresponds to a part of the somatosensory cortex of an adult mouse and was acquired using serial section electron microscopy (ssEM). The train and test volume dimensions are $85\times1463\times1613\:\mbox{voxels}$ and $75\times1334\times1553\:\mbox{voxels}$ respectively, with an anisotropic resolution of $3\times3\times30$ nm per voxel. An image sample is presented in Figure \ref{fig:datasets} (blue).

\subsection{Experimental setup}

\textbf{Evaluation metrics.}
\label{sec:eval_metrics}
Following common practice in the field, we evaluate our methods using the Jaccard index of the positive class or \textit{foreground IoU}.  

The foreground IoU is defined as follows: 

\begin{equation}
\label{jaccardequation}
    IoU_{F}(A, B) = \frac{|A \cap B|}{|A \cup B|} = \frac{TP}{TP+FP+FN}
\end{equation}

where $A$ is the segmentation proposal, $B$ is the ground truth, TP are the true positives, FP the false positives and FN are the false negatives. As a convention, the positive class is foreground and the negative class is background. The background IoU is defined likewise by swapping the positive and negative classes. To obtain these values, the probability image returned by the network is binarized using a threshold value of $0.5$.

Nevertheless, to compare our results with other related works we also define the \textit{overall IoU} as the average of the foreground and background IoU:

\begin{equation}
    \label{VOCequation}
    IoU_{O} = \frac{(IoU_{F} + IoU_{B})}{2}
\end{equation}

where $IoU_{F}$ and $IoU_{B}$ are the foreground and background IoU, respectively. Notice the high proportion of background pixels typically inflates the overall IoU score, resulting in greater values than the foreground IoU. 

\textbf{Training setup and data augmentation.} In order to find the best solutions, we made an exhaustive search of hyperparameters and training configurations, exploring different loss functions, optimizers, learning rates, batch sizes and data augmentation techniques. We explored as well the use of different input patch sizes, their selection method (random or systematic), and the discarding of image patches with low foreground class information as in~\cite{oztel2017mitochondria}. When selecting a random patch, we define a probability map to choose patches with a higher probability of containing mitochondria, therefore addressing the class imbalance problem. Finally, we have also studied the effect of selecting the validation set as either consecutive training images or at random. Here we describe the best training configuration found. However, the details of this exhaustive search of the best training workflow can be consulted in the appendices. 

In particular, we minimize the binary cross entropy (BCE) loss using the Stochastic Gradient Descent (SGD) optimizer, $0.99$ momentum and no decay, with a learning rate of $0.002$ and a batch size value of $6$. The validation set is formed by $10\%$ of the training images selected at random. We use a GeForce GTX 1080 GPU card to train the network for $360$ epochs, completing an epoch when all training data is explored, with a patience established at $100$ epochs monitoring the validation loss and picking up the model that performs best in the validation set. Moreover, we apply data augmentation on-the-fly making random rotations and vertical and horizontal flips. For the 3D networks, we employ elastic transformations as well (in 2D we did not observe an improvement).

\subsection{Experiments on Lucchi dataset}

\subsubsection{Reproducing top state-of-the-art methods}
\label{reproduce_sota_works}


In this section, we present a clear example of the difficulty of reproducing published methods when the training workflow is not fully described and the code is not provided. In particular, we aimed at reproducing the state-of-the-art deep learning-based methods that report top performance in the Lucchi dataset published by Cheng \etal~\cite{cheng2017volume}, Casser \etal~\cite{casser2020fast}, Xiao \etal~\cite{xiao2018automatic} and Oztel \etal~\cite{oztel2017mitochondria}. Only the code by Casser \etal~\cite{casser2020fast} is publicly available, so we plugged their network architecture into our training workflow. The code from the rest of methods was unsuccessfully requested to their corresponding authors.

In all cases, a first implementation attempt was made following the methodology and exact parameters described on each publication. When finding missing information, we proceeded using the most common practice in the field. On top of that, following the same procedure we use for our own models, we modified the original configuration (i.e., architecture and training workflow) aiming at improving  the results and their stability (full details available in the appendices). These configurations are hereafter referred to as \textit{original} and \textit{modified} respectively. As can be seen in Table~\ref{table:sota_reproduction}, although a systematic search of the best hyperparameters and training configurations was performed, the scores obtained in some cases do not reach the reported ones.

The original configuration of the 2D network presented by Cheng \etal~\cite{cheng2017volume} produces very unstable results (high standard deviation), probably due to the high learning rate employed ($0.05$), even though it is reduced when reaching the $50\%$ and $75\%$ of total epochs. Our modified configuration differs in the optimizer used (Adam instead of SGD) and learning rate (fixed to $0.0001$). Additionally, we performed extra DA with random rotations, removed the dropout layers, reduced the number of epochs and extracted 12 random patches per training image instead of just one. Without post-processing (none is used in the original publication), the foreground IoU value reported ($0.865$) can only be reached through our modified configuration and by taking the maximum values of the \textit{$50\%$ overlap} or \textit{full image} reconstruction strategies. Notice that even better values can be obtained thanks to post-processing.

The 3D approach of the same authors~\cite{cheng2017volume} produces IoU values close to 0 in its original form, since using the proposed learning rate ($0.1$), the network gets easily trapped in local minima. Moreover, the subvolume shape adopted, $128\times128\times96\ \mbox{pixels}$, makes train/validation data splitting difficult, so we train the network until convergence with no validation data. Our modified configuration produces better results but far from the reported ones and highly unstable ($0.800$ in its best run vs the reported $0.889$).

The original configuration proposed by Casser \etal~\cite{casser2020fast} reaches high IoU values with high standard deviation as well. We modified their configuration selecting two random patches per training image instead of one and using a probability map to prioritize patches having mitochondria pixels in the center, which leads to more stable results. The maximum value was obtained applying Z-filtering to the predictions over full test images, measuring $0.870$ of foreground IoU. In their original code, Casser \etal~\cite{casser2020fast} optimized the training by using the test set as validation set, what could explain their better reported value.

{\centering\begin{sideways}
    \begin{minipage}{1.1\textheight}
        \resizebox{1.1\textheight}{!}{%
            \setstretch{1.5}
            \begin{tabular}{rrrrrrrrrrrrrrrrrr}
\hline
\textbf{}                          & \textbf{}                                                                    & \textbf{}          & \phantom{abc} & \textbf{}               & \phantom{abc} & \multicolumn{2}{c}{\textbf{Per Image (no overlap)}}            & \phantom{abc} & \multicolumn{5}{c}{\textbf{Per image (50\% overlap)}}                                                                                                                                                                                                                              & \phantom{abc} & \multicolumn{3}{c}{\textbf{Full Image}}                                                                     \\ \cline{7-8} \cline{10-14} \cline{16-18} 
\textbf{Network}                   & \textbf{\begin{tabular}[c]{@{}r@{}}Param.\\ number\end{tabular}} & \textbf{~Reported} & \textbf{}     & \textbf{Per Patch}      & \textbf{}     & \textbf{}               & \textbf{+Z-Fil.}        & \textbf{}     & \textbf{}       & \textbf{+Ensemble} & \textbf{\begin{tabular}[c]{@{}r@{}}+Ensemble\\ +Z-Fil.\end{tabular}} & \textbf{\begin{tabular}[c]{@{}r@{}}+Blended \\ Ensemble\end{tabular}} & \textbf{\begin{tabular}[c]{@{}r@{}}+Blended \\ Ensemble\\ +Z. Fil.\end{tabular}} & \textbf{}     & \textbf{}       & \textbf{+Ensemble} & \textbf{\begin{tabular}[c]{@{}r@{}}+Ensemble\\ +Z-Fil.\end{tabular}} \\ \hline
Cheng 2D~\cite{cheng2017volume}    & 0.6M                                                                          & 0.865              &               &                         &               &                         &                         &               &                 &                    &                                                                      &                                                                       &                                                                                  &               &                 &                    &                                                                      \\
\textit{Original}                                 & 0.59M                                                                         &                    &               & 0.503$\pm$0.233         &               & 0.503$\pm$0.233         & 0.511$\pm$0.238         &               & 0.517$\pm$0.240 & 0.517$\pm$0.239    & 0.521$\pm$0.243                                                      & 0.541$\pm$0.250                                                       & 0.548$\pm$0.254                                                                  &               & 0.526$\pm$0.244 & 0.537$\pm$0.244    & 0.543$\pm$0.252                                                      \\
\textit{Modified}                                 & 0.59M                                                                         &                    &               & 0.848$\pm$0.012         &               & 0.843$\pm$0.012         & 0.852$\pm$0.011         &               & 0.851$\pm$0.011 & 0.863$\pm$0.010    & 0.868$\pm$0.010                                                      & 0.865$\pm$0.008                                                       & 0.871$\pm$0.008                                                                  &               & 0.853$\pm$0.011 & 0.865$\pm$0.009    & 0.871$\pm$0.008                                                      \\
Maximum                            & -                                                                            &                    &               & 0.864                   &               & 0.858                   & 0.865                   &               & 0.865           & 0.877              & 0.881                                                                & 0.878                                                                 & 0.883                                                                            &               & 0.865           & 0.878              & 0.881                                                                \\
                                   &                                                                              &                    &               &                         &               &                         &                         &               &                 &                    &                                                                      &                                                                       &                                                                                  &               &                 &                    &                                                                      \\
Casser~\cite{casser2020fast}       & 1.96M                                                                         & 0.890              &               &                         &               &                         &                         &               &                 &                    &                                                                      &                                                                       &                                                                                  &               &                 &                    &                                                                      \\
\textit{Original}                                 & 1.96M                                                                         &                    &               & 0.824$\pm$0.014         &               & 0.817$\pm$0.016         & 0.828$\pm$0.016         &               & 0.815$\pm$0.016 & 0.825$\pm$0.013    & 0.831$\pm$0.013                                                      & 0.831$\pm$0.011                                                       & 0.838$\pm$0.011                                                                  &               & 0.820$\pm$0.016 & 0.833$\pm$0.011    & 0.839$\pm$0.012                                                      \\
\textit{Modified}                                 & 1.96M                                                                         &                    &               & 0.844$\pm$0.014         &               & 0.838$\pm$0.008         & 0.845$\pm$0.009         &               & 0.837$\pm$0.008 & 0.846$\pm$0.016    & 0.850$\pm$0.017                                                      & 0.850$\pm$0.016                                                       & 0.855$\pm$0.017                                                                  &               & 0.842$\pm$0.006 & 0.853$\pm$0.015    & 0.858$\pm$0.015                                                      \\
Maximum                            & -                                                                            &                    &               & 0.846                   &               & 0.844                   & 0.852                   &               & 0.846           & 0.861              & 0.865                                                                & 0.862                                                                 & 0.867                                                                            &               & 0.848           & 0.865              & 0.870                                                                \\
                                   &                                                                              &                    &               &                         &               &                         &                         &               &                 &                    &                                                                      &                                                                       &                                                                                  &               &                 &                    &                                                                      \\
Oztel~\cite{oztel2017mitochondria} & 0.14M                                                                         & 0.907              &               &                         &               &                         &                         &               &                 &                    &                                                                      &                                                                       &                                                                                  &               &                 &                    &                                                                      \\
\textit{Original}                                 & 0.14M                                                                         &                    &               & -                       &               & -                       & -                       &               & -               & -                  & -                                                                    & -                                                                     & -                                                                                &               & 0.425$\pm$0.080 & 0.457$\pm$0.060    & 0.466$\pm$0.061                                                      \\
\textit{Modified}                                 & 0.07M                                                                         &                    &               &                         &               & -                       & -                       &               & -               & -                  & -                                                                    & -                                                                     & -                                                                                &               & 0.451$\pm$0.042 & 0.476$\pm$0.049    & 0.487$\pm$0.053                                                      \\
Maximum                            & -                                                                            &                    &               & -                       &               & -                       & -                       &               & -               & -                  & -                                                                    & -                                                                     & -                                                                                &               & 0.500           & 0.531              & 0.544                                                                \\
                                   &                                                                              &                    &               &                         &               &                         &                         &               &                 &                    &                                                                      &                                                                       &                                                                                  &               &                 &                    &                                                                      \\ \hline
Cheng 3D~\cite{cheng2017volume}    & 0.63M                                                                         & 0.889              &               &                         &               &                         &                         &               &                 &                    &                                                                      &                                                                       &                                                                                  &               &                 &                    &                                                                      \\
\textit{Original}                                 & 0.79M                                                                         &                    &               & 0.053$\pm$0.000$(\dag)$ &               & 0.053$\pm$0.000$(\dag)$ & 0.053$\pm$0.000$(\dag)$ &               & 0.053$\pm$0.000 & 0.053$\pm$0.000    & 0.053$\pm$0.000                                                      & -                                                                     & -                                                                                &               & -               & -                  & -                                                                    \\
\textit{Modified}                                 & 0.79M                                                                         &                    &               & 0.623$\pm$0.039$(\dag)$ &               & 0.691$\pm$0.049$(\dag)$ & 0.693$\pm$0.049$(\dag)$ &               & 0.714$\pm$0.040 & 0.0737$\pm$0.034    & 0.738$\pm$0.034                                                      & -                                                                     & -                                                                                &               & -               & -                  & -                                                                    \\
Maximum                            & -                                                                            &                    &               & 0.694                   &               & 0.777                   & 0.779                   &               & 0.787           & 0.799              & 0.800                                                                & -                                                                     & -                                                                                &               & -               & -                  & -                                                                    \\
                                   &                                                                              &                    &               &                         &               &                         &                         &               &                 &                    &                                                                      &                                                                       &                                                                                  &               &                 &                    &                                                                      \\
Xiao~\cite{xiao2018automatic}      & 1.1M                                                                          & 0.900              &               &                         &               &                         &                         &               &                 &                    &                                                                      &                                                                       &                                                                                  &               &                 &                    &                                                                      \\
\textit{Original}                                 & 1.08M                                                                         &                    &               & 0.874$\pm$0.003$(\dag)$ &               & 0.863$\pm$0.004$(\dag)$ & 0.864$\pm$0.004$(\dag)$ &               & 0.863$\pm$0.004 & 0.866$\pm$0.004    & 0.867$\pm$0.004                                                      & -                                                                     & -                                                                                &               & -               & -                  & -                                                                    \\
\textit{Modified}                                 & 1.08M                                                                         &                    &               & 0.882$\pm$0.002$(\dag)$ &               & 0.873$\pm$0.003$(\dag)$ & 0.874$\pm$0.003$(\dag)$ &               & 0.872$\pm$0.003 & 0.874$\pm$0.003    & 0.874$\pm$0.003                                                      & -                                                                     & -                                                                                &               & -               & -                  & -                                                                    \\
Maximum                            & -                                                                            &                    &               & 0.885                   &               & 0.879                   & 0.880                   &               & 0.880           & 0.880              & 0.880                                                                & -                                                                     & -                                                                                &               & -               & -                  & -                                                                    \\ \hline
\end{tabular}
        }
        \captionof{table}{Foreground IoU (mean$\pm$standard deviation) of reproduced state-of-the-art works in Lucchi dataset. Different scores discussed in previous sections are shown, the post-processing methods adopted are indicated (\textit{Z-Fil.} refers to Z-filtering). \textit{Blended Ensemble} refers to combining blending and ensemble estimation. \textit{Original} versions refer to exact configurations as reported by the authors. \textit{Modified} corresponds to our best approach modifying \textit{Original} in some way to improve method's performance and results stability. The patch size and overlap (marked with $\dag$) in each work is as follows: $256\times256\:\mbox{pixels}$ for Cheng 2D, $128\times128\times96\:\mbox{voxels}$ ($0\times0\times27\:\mbox{voxels}$ overlap in $x\times y\times z$) for the subvolumes in Cheng 3D; $512\times512\:\mbox{pixels}$ ($256\times0\:\mbox{pixels}$ overlap in $x\times y$) in Casser; $448\times576\times20\:\mbox{voxels}$ ($128\times128\times10\:\mbox{voxels}$ overlap in $x\times y\times z$) in Xiao; and $768\times1024$ pixels in Oztel.}
        \label{table:sota_reproduction}
    \end{minipage}
\end{sideways}\par}

Xiao \etal~\cite{xiao2018automatic} provided a detailed explanation of their training procedure, architecture and output reconstruction strategy. Thus, the unique modification that we added is the use of elastic transformations in DA. As it is shown in Table $1$, this change improves substantially the results obtained. They merge the predictions with overlap and ensemble, so to be fair, the maximum value of patch merging using $50\%$ overlap and ensemble predictions should be used for comparison. They reported $0.900$ of foreground IoU compared to the maximum $0.880$ achieved by our modified version.

Finally, the original configuration proposed by Oztel \textit{et al.}~\cite{oztel2017mitochondria} produces very low foreground IoU values. Indeed, the amount of relevant details regarding the architecture, hyperparameter and post-processing methods that are missing in the original publication is remarkable. Thus, even after modifying their network by adding more non-linearities (ReLU), changing the dropout values or the feature maps used, the results obtained are far from those presented by the authors. The number of the original network parameters compared with other state-of-the-art approaches is also relatively low, only $0.14$M, which in our opinion are insufficient to capture mitochondria shapes. Furthermore, we implemented their post-processing pipeline, whose results are presented in Table~\ref{table:oztel_post}. We adapted these post-processing methods to specifically improve the segmentation made by the proposed network. Although the segmentation is improved by a large margin, it is still far to achieve the IoU reported by the authors.

\begin{table}[]
\ra{1.3}
\centering
\begin{tabular}{lcccccccc}
\hline
        & \phantom{abc}        & \textbf{\begin{tabular}[c]{@{}c@{}}Full\\ Image\end{tabular}} & \phantom{abc}        & \textbf{\begin{tabular}[c]{@{}c@{}}Spurious \\ Detection\end{tabular}} & \phantom{abc}        & \textbf{Watershed}                  & \phantom{abc}        & \textbf{Z-Filtering}                \\ \hline
\textit{Original}      & \multicolumn{1}{r}{} & \multicolumn{1}{r}{0.425$\pm$0.080}                           & \multicolumn{1}{r}{} & \multicolumn{1}{r}{0.426$\pm$0.091}                                    & \multicolumn{1}{r}{} & \multicolumn{1}{r}{0.540$\pm$0.100} & \multicolumn{1}{r}{} & \multicolumn{1}{r}{0.573$\pm$0.106} \\
\textit{Modified}      &                      & \multicolumn{1}{r}{0.451$\pm$0.042}                           &                      & \multicolumn{1}{r}{0.449$\pm$0.067}                                    &                      & \multicolumn{1}{r}{0.562$\pm$0.057} &                      & \multicolumn{1}{r}{0.599$\pm$0.067} \\
Maximum &                      & 0.500                                                         &                      & 0.539                                                                  &                      & 0.619                               &                      & 0.683                               \\ \hline
\end{tabular}
\caption{Foreground IoU results by the original and modified configurations of Oztel \etal~\cite{oztel2017mitochondria}  using their consecutive post-processing methods, i.e., \textit{Spurious Detection} is applied over \textit{Full Images}, then they are passed through \textit{Watershed}, and finally through Z-filtering.}
\label{table:oztel_post}
\end{table}

\subsubsection{Proposed networks vs state-of-the-art networks for semantic segmentation}
\label{sec:other_sota_networks}

In this section, we introduce the performance of our proposed architectures together with a study in-depth of the main state-of-the-art semantic segmentation networks for natural and biomedical images. Namely, FCN 8/32~\cite{long2015fully}, MultiResUNet~\cite{ibtehaz2020multiresunet}, MNet~\cite{fu2018joint}, Tiramisu~\cite{jegou2017one}, U-Net++~\cite{zhou2018unet++}, 3D Vanilla U-Net~\cite{cciccek20163d} and nnU-Net~\cite{isensee2021nnu}. All implementations have been obtained or ported from their official sites and all networks have been optimized under the same conditions: same training and validation partitions, DA transformations, optimizers and learning rate ranges (see appendices). The case of the nnU-Net is special, since it is designed to optimize the whole segmentation pipeline. In order to compare it in equal conditions with the other approaches, we extract the optimal architecture following the nnU-Net regular processing and plugged it into our own workflow.

For all 2D networks, we use an input patch size of $256\times256$ pixels, while for the 3D networks we use $80\times80\times80\;\mbox{voxels}$ subvolumes to exploit the isotropic resolution of the Lucchi dataset. Notice the difference between the input shapes of 2D and 3D architectures makes them not directly comparable when looking at the metric values obtained per patch or with a $0\%$ overlap output reconstruction, since some overlap was needed in 3D to infer the whole test volume. Therefore, for a fair comparison between 2D and 3D networks, we refer to the results based on output reconstructions using $50\%$ of patch overlap. 

The results from the best configuration found for each network are shown in Table \ref{table:other_sota_networks}. Notice the 3D networks do not have results using full image reconstructions due to GPU memory limitations, as the whole dataset should be fed to the network. In the same way, blending estimation was not implemented in 3D networks given their computational cost.

\textbf{Performance of state-of-the-art biomedical segmentation networks.} The results of Tiramisu~\cite{jegou2017one}, MNet~\cite{fu2018joint}, nnU-Net~\cite{isensee2021nnu}, MultiResUNet~\cite{ibtehaz2020multiresunet} and 3D Vanilla U-Net~\cite{cciccek20163d} are below $0.880$ of foreground IoU even when using output reconstructions with 50\% of overlap and post-processing techniques such as ensemble predictions or Z-filtering. On top of these networks, the U-Net++ achieved the best results, scoring $0.881\pm 0.004$ of foreground IoU. The 3D Vanilla U-Net, nnU-Net, U-Net++ and MNet seem to produce stable results (low standard deviation), while Tiramisu and MultiResUNet have larger variability within their results. Besides that, the difference in their number of trainable parameters is remarkable. The 3D Vanilla U-Net, nnU-Net and U-Net++ models have between $2\times$ and $5\times$ more parameters than the other state-of-the-art approaches. 

With respect to the FCN networks~\cite{long2015fully}, the FCN32 reports low IoU values that do not reach state-of-the-art results, but the FCN8 achieves results comparable with our best 2D U-Net configuration. Nevertheless, the number of trainable parameters in FCN8 is $50.38$M compared to less than 2M in our proposed 2D models. 

{\centering\begin{sideways}
    \begin{minipage}{1.13\textheight}
        \resizebox{1.13\textheight}{!}{%
            \setstretch{1.5}
\begin{tabular}{rrlrrrrrrrrrrrrrr}
\hline
\multicolumn{1}{c}{\textbf{}}               & \multicolumn{1}{c}{\textbf{}}                                    & \multicolumn{1}{c}{\phantom{abc}} & \multicolumn{1}{c}{\textbf{}}                                 & \multicolumn{1}{c}{\phantom{abc}} & \multicolumn{2}{c}{\textbf{Per image (no overlap)}}             & \multicolumn{1}{c}{\phantom{abc}} & \multicolumn{5}{c}{\textbf{Per image (50\% overlap)}}                                                                                                                                                                                                                                                               & \multicolumn{1}{c}{\phantom{abc}} & \multicolumn{3}{c}{\textbf{Full Image}}                                                                                    \\ \cline{6-7} \cline{9-13} \cline{15-17} 
\textbf{Network}                            & \textbf{\begin{tabular}[c]{@{}r@{}}Param.\\ number\end{tabular}} &                                   & \textbf{\begin{tabular}[c]{@{}r@{}}Per \\ Patch\end{tabular}} &                                   & \textbf{}                        & \textbf{+Z-Fil.}                 &                                   & \textbf{}                & \textbf{+Ensemble}       & \textbf{\begin{tabular}[c]{@{}r@{}}+Ensemble\\ +Z-Fil.\end{tabular}} & \textbf{\begin{tabular}[c]{@{}r@{}}+Blended\\ Ensemble\end{tabular}} & \multicolumn{1}{c}{\textbf{\begin{tabular}[c]{@{}c@{}}+Blended\\ Ensemble\\ + Z-Fil.\end{tabular}}} &                                   & \textbf{}                & \textbf{+Ensemble}       & \textbf{\begin{tabular}[c]{@{}r@{}}+Ensemble\\ +Z-Fil.\end{tabular}} \\ \hline
FCN 32~\cite{dai2016r}                      & 50.38M                                                           &                                   & 0.040$\pm$0.000                                               &                                   & 0.637$\pm$0.005                  & 0.640$\pm$0.005                  &                                   & 0.677$\pm$0.005          & 0.679$\pm$0.006          & 0.680$\pm$0.006                                                      & 0.659$\pm$0.004                                                      & 0.661$\pm$0.004                                                                                     &                                   & 0.657$\pm$0.003          & 0.659$\pm$0.003          & 0.660$\pm$0.003                                                      \\

MultiResUNet~\cite{ibtehaz2020multiresunet} & 7.26M                                                            &                                   & 0.815$\pm$0.000                                               &                                   & 0.812$\pm$0.016                  & 0.821$\pm$0.015                  &                                   & 0.814$\pm$0.014          & 0.820$\pm$0.010          & 0.824$\pm$0.010                                                      & 0.834$\pm$0.010                                                      & 0.840$\pm$0.009                                                                                     &                                   & 0.828$\pm$0.016          & 0.833$\pm$0.010          & 0.839$\pm$0.010                                                      \\
Tiramisu~\cite{jegou2017one}                & 9.4M                                                             &                                   & 0.810$\pm$0.028                                               &                                   & 0.809$\pm$0.030                  & 0.821$\pm$0.029                  &                                   & 0.833$\pm$0.027          & 0.851$\pm$0.018          & 0.857$\pm$0.017                                                      & 0.850$\pm$0.016                                                      & 0.855$\pm$0.016                                                                                     &                                   & 0.830$\pm$0.029          & 0.846$\pm$0.019          & 0.851$\pm$0.018                                                      \\
MNet~\cite{fu2018joint}                     & 8.54M                                                            &                                   & 0.851$\pm$0.011                                               &                                   & 0.854$\pm$0.009                  & 0.861$\pm$0.009                  &                                   & 0.865$\pm$0.008          & 0.870$\pm$0.007          & 0.874$\pm$0.007                                                      & 0.874$\pm$0.006                                                      & 0.878$\pm$0.006                                                                                     &                                   & 0.867$\pm$0.008          & 0.872$\pm$0.006          & 0.876$\pm$0.008                                                      \\
nnU-Net~\cite{isensee2021nnu}               & 52.1M                                                            & \multicolumn{1}{r}{}              & 0.845$\pm$0.009                                               &                                   & 0.845$\pm$0.009                  & 0.853$\pm$0.010                  &                                   & 0.854$\pm$0.011          & 0.872$\pm$0.005          & 0.876$\pm$0.006                                                      & 0.876$\pm$0.006                                                      & 0.881$\pm$0.005                                                                                     &                                   & 0.799$\pm$0.052          & 0.788$\pm$0.066          & 0.790$\pm$0.068                                                      \\
U-Net++~\cite{zhou2018unet++}               & 37.7M                                                            & \multicolumn{1}{r}{}              & 0.731$\pm$0.014                                               &                                   & 0.860$\pm$0.008                  & 0.867$\pm$0.008                  &                                   & 0.872$\pm$0.005          & 0.877$\pm$0.004          & 0.881$\pm$0.004                                                      & 0.880$\pm$0.003                                                      & 0.884$\pm$0.003                                                                                     &                                   & 0.875$\pm$0.004          & 0.878$\pm$0.003          & 0.882$\pm$0.003                                                      \\
2D Residual U-Net (ours)                    & 2.03M                                                            &                                   & 0.867$\pm$0.005                                               &                                   & 0.864$\pm$0.005                  & 0.871$\pm$0.006                  &                                   & 0.873$\pm$0.005          & 0.877$\pm$0.004          & 0.880$\pm$0.004                                                      & 0.878$\pm$0.003                                                      & 0.882$\pm$0.003                                                                                     &                                   & 0.875$\pm$0.004          & 0.877$\pm$0.003          & 0.880$\pm$0.004                                                      \\

2D SE U-Net (ours)                          & 1.95M                                                            &                                   & 0.863$\pm$0.002                                               &                                   & 0.861$\pm$0.003                  & 0.869$\pm$0.003                  &                                   & 0.873$\pm$0.003          & 0.878$\pm$0.003          & 0.882$\pm$0.003                                                      & 0.880$\pm$0.003                                                      & 0.883$\pm$0.003                                                                                     &                                   & 0.875$\pm$0.002          & 0.881$\pm$0.002          & 0.881$\pm$0.002                                                      \\

FCN 8~\cite{dai2016r}                       & 50.38M                                                           &                                   & 0.860$\pm$0.005                                               &                                   & 0.864$\pm$0.005                  & 0.871$\pm$0.005                  &                                   & 0.880$\pm$0.003          & 0.884$\pm$0.002          & 0.888$\pm$0.002                                                      & \textbf{0.887$\pm$0.002}                                             & 0.891$\pm$0.002                                                                                     &                                   & 0.881$\pm$0.003          & 0.886$\pm$0.002          & \textbf{0.891$\pm$0.002}                                             \\

2D U-Net (ours)                             & \textbf{1.95M}                                                   &                                   & 0.874$\pm$0.003                                               &                                   & 0.872$\pm$0.003                  & 0.880$\pm$0.003                  &                                   & 0.881$\pm$0.002          & 0.884$\pm$0.002          & 0.888$\pm$0.002                                                      & 0.884$\pm$0.000                                                      & 0.889$\pm$0.002                                                                                     &                                   & 0.882$\pm$0.003          & 0.884$\pm$0.002          & 0.887$\pm$0.003                                                      \\
2D Attention U-Net (ours)                   & 1.99M                                                            &                                   & \textbf{0.875$\pm$0.004}                                      &                                   & \textbf{0.873$\pm$0.003}         & \textbf{0.882$\pm$0.003}         &                                   & \textbf{0.882$\pm$0.003} & \textbf{0.885$\pm$0.001} & \textbf{0.890$\pm$0.002}                                             & 0.886$\pm$0.001                                                      & \textbf{0.892$\pm$0.001}                                                                            &                                   & \textbf{0.884$\pm$0.002} & \textbf{0.886$\pm$0.001} & 0.890$\pm$0.002                                                      \\ \hline
3D Vanilla U-Net~\cite{cciccek20163d}       & 19.07M                                                           &                                   & 0.402$\pm$0.005$(\dag)$                                       &                                   & 0.842$\pm$0.004$(\dag)$          & 0.844$\pm$0.005$(\dag)$          &                                   & 0.851$\pm$0.004          & 0.857$\pm$0.006          & 0.857$\pm$0.006                                                      & -                                                                    & -                                                                                                   &                                   & -                        & -                        & -                                                                    \\
3D SE U-Net (ours)                          & 0.79M                                                            &                                   & 0.387$\pm$0.007$(\dag)$                                       &                                   & 0.854$\pm$0.013$(\dag)$          & 0.855$\pm$0.013$(\dag)$          &                                   & 0.867$\pm$0.009          & 0.873$\pm$0.007          & 0.874$\pm$0.007                                                      & -                                                                    & -                                                                                                   &                                   & -                        & -                        & -                                                                    \\
3D Attention U-Net (ours)                   & 0.79M                                                            &                                   & 0.389$\pm$0.005$(\dag)$                                       &                                   & 0.856$\pm$0.003$(\dag)$          & 0.857$\pm$0.003$(\dag)$          &                                   & 0.870$\pm$0.003          & 0.876$\pm$0.003          & 0.876$\pm$0.003                                                      & -                                                                    & -                                                                                                   &                                   & -                        & -                        & -                                                                    \\
3D U-Net (ours)                             & \textbf{0.79M}                                                   &                                   & 0.394$\pm$0.005$(\dag)$                                       &                                   & \textbf{0.858$\pm$0.007$(\dag)$} & \textbf{0.859$\pm$0.007$(\dag)$} &                                   & 0.871$\pm$0.006          & 0.878$\pm$0.004          & 0.878$\pm$0.004                                                      & -                                                                    & -                                                                                                   &                                   & -                        & -                        & -                                                                    \\
3D Residual U-Net (ours)                    & 1.50M                                                            &                                   & \textbf{0.394$\pm$0.004$(\dag)$}                              &                                   & 0.857$\pm$0.004$(\dag)$          & 0.858$\pm$0.004$(\dag)$          &                                   & \textbf{0.877$\pm$0.004} & \textbf{0.883$\pm$0.002} & \textbf{0.883$\pm$0.002}                                             & -                                                                    & -                                                                                                   &                                   & -                        & -                        & -                                                                    \\ \hline
\end{tabular}

}
        \captionof{table}{Performance of proposed networks and state-of-the-art networks for semantic segmentation in the Lucchi dataset. All values represent the foreground IoU (mean$\pm$standard deviation). Scores are shown using the different post-processing methods adopted (\textit{Z-Fil.} refers to Z-filtering). \textit{Blended ensemble} refers to combining blending and ensemble estimation. In 3D patches a minimum overlap was required so they are marked with $\dag$. Best results of each column and type of network (2D or 3D) are shown in bold.}
\label{table:other_sota_networks}
    \end{minipage}
\end{sideways}\par}

\begin{figure}[]
\begin{subfigure}{0.5\textwidth}
  \includegraphics[width=\linewidth]{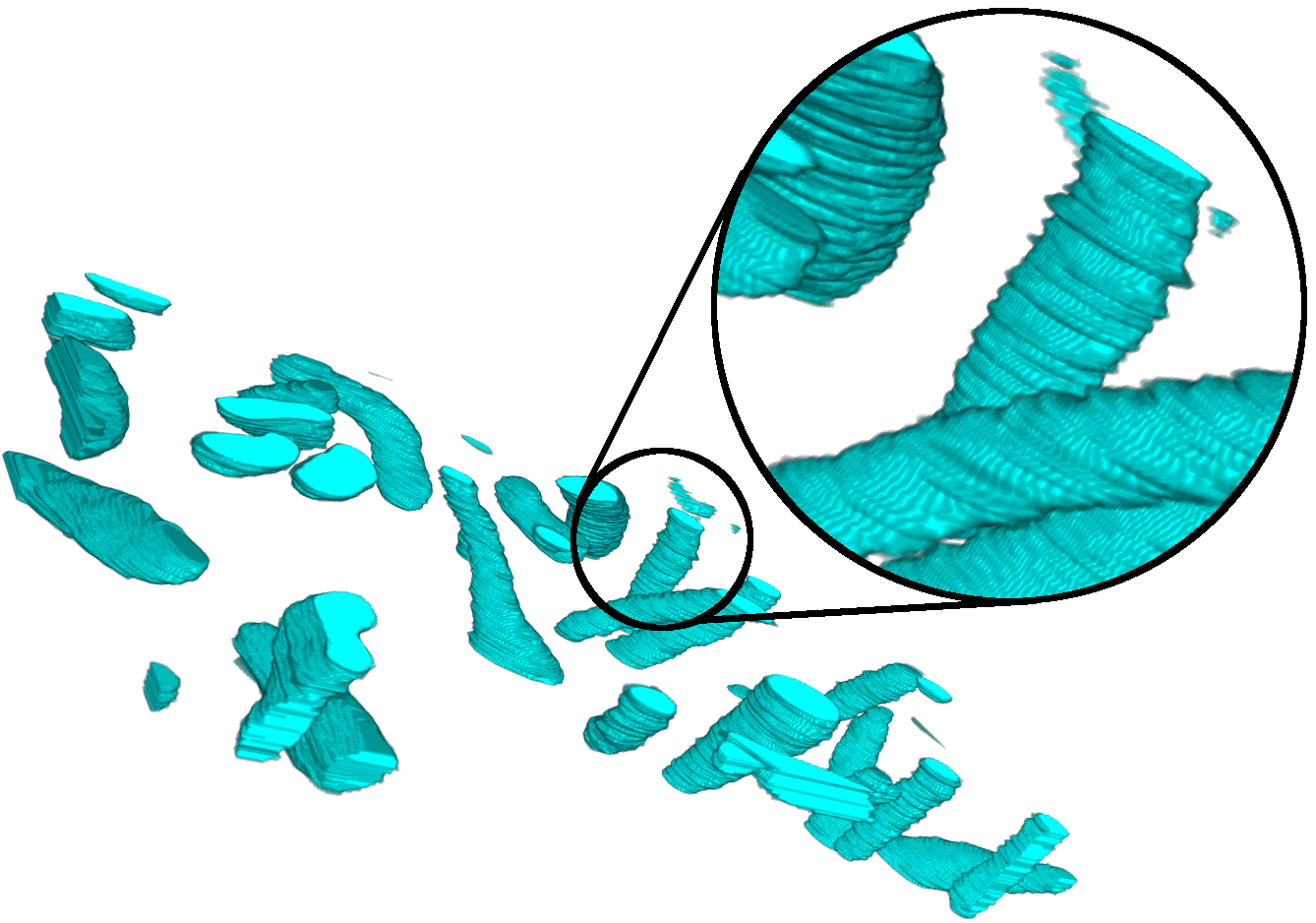}
  \caption{ }
  \label{fig:lucchi_lucchi++_comp_a}
\end{subfigure}\hfil 
\begin{subfigure}{0.5\textwidth}
  \includegraphics[width=\linewidth]{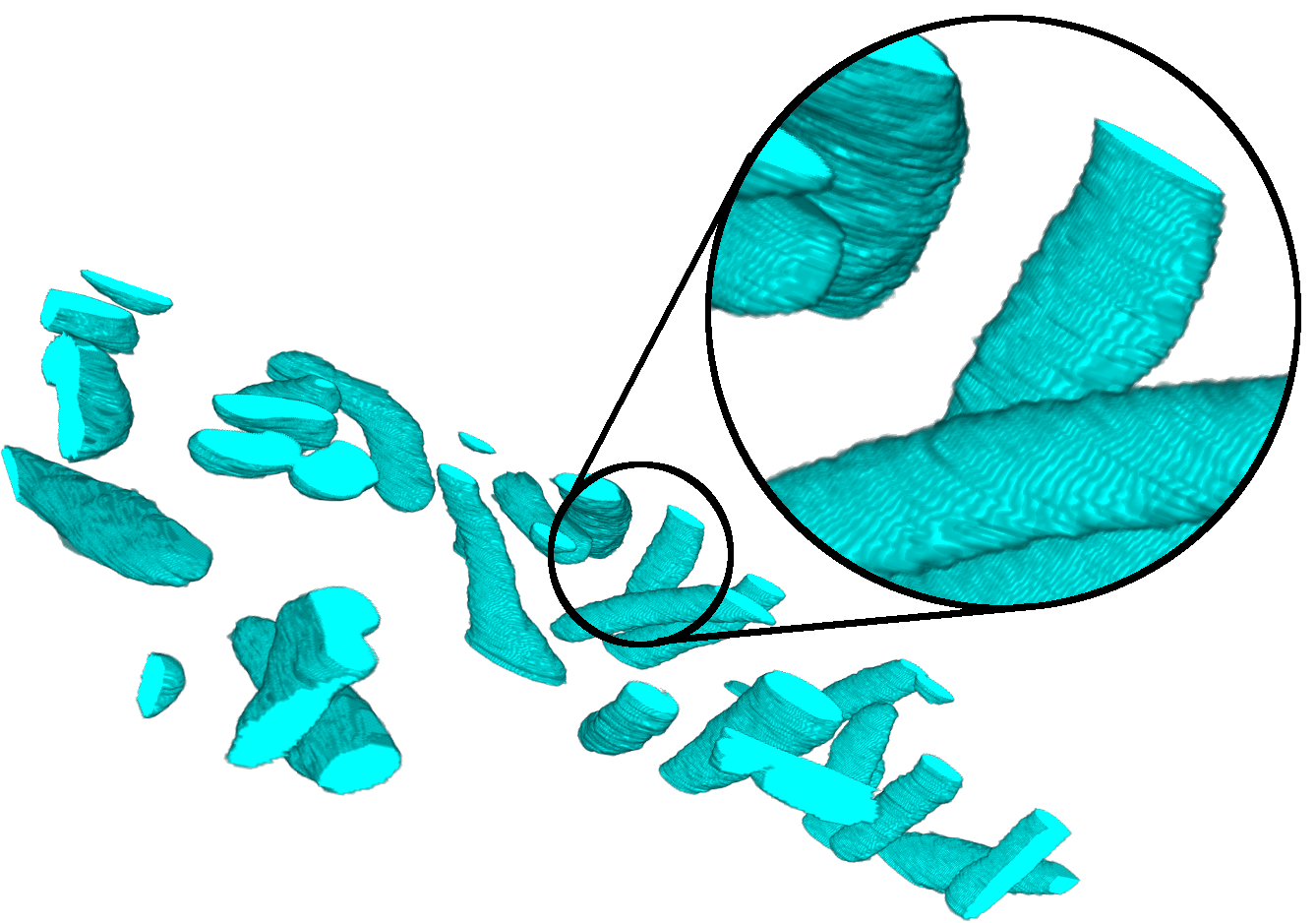}
  \caption{ }
  \label{fig:lucchi_lucchi++_comp_b}
\end{subfigure}
\caption{3D view of mitochondria labels on: (a) Lucchi and (b) Lucchi++ datasets. Labels in Lucchi are more jagged than in Lucchi++, penalizing 3D networks performance.}
\end{figure}

\textbf{Performance of our proposed networks.} Regarding our proposed approaches (Section \ref{sec:proposed_networks}), the best values were obtained with the 2D U-Net and its version with attention gates: $0.888\pm 0.002$ and $0.890\pm 0.002$ respectively applying ensemble estimation and Z-filtering post-processing. Our 3D networks do not reach the performance obtained with 2D versions. This may be explained by, first, the difference between the number of data samples. In 2D, with patches of $256\times256$ pixels, the training set has $1980$ samples, while in 3D, with a subvolume size of $80\times80\times80$ voxels, the number of training samples without overlapping areas results in $216$. Nevertheless, we try to alleviate this gap by means of elastic transformations in DA. Second, while inspecting mitochondria labels in 3D, we observed they frequently lose shape continuity through slices, penalizing the learning capacity of 3D networks (see Figure \ref{fig:lucchi_lucchi++_comp_a}). 
Remarkably, our 3D networks have three times less training parameters than our 2D approaches, leading to more computationally efficient models. 

Finally, to complete the overview of the state-of-the-art networks and architectures, we experimented with  \textit{Squeeze-and-Excitation} (SE) blocks~\cite{hu2018squeeze} in our proposed 2D and 3D models. 
These blocks perform dynamic channel-wise feature recalibration by \textit{squeeze} and \textit{excite} operations. The \textit{Squeeze} operation consists in collecting global spatial information into a channel descriptor using global average pooling. After that, features are recalibrated by the \textit{excite} operation, which emphasizes channel-wise features with a simple gating mechanism based on a ReLU and a Sigmoid activation. Their best results are obtained with SE blocks everywhere except the bottleneck, as suggested by~\cite{roy2018recalibrating}. Nevertheless, we experimented as well with inserting SE blocks after every convolutional layer. As shown in Table \ref{table:other_sota_networks}, these blocks do not imply a boost in performance in this case.

{\centering\begin{sideways}
    \begin{minipage}{1\textheight}
        \resizebox{1\textheight}{!}{%
            \setstretch{1.2}

\begin{tabular}{rlclclclccclc}
\hline
\textbf{}                        & \phantom{abc} & \textbf{}                                   & \phantom{abc} & \textbf{} & \phantom{abc}    & \multicolumn{3}{c}{\textbf{Foreground IoU}}                                                                                                             & \phantom{abc} & \multicolumn{3}{c}{\textbf{Overall IoU}}                                                                                     \\ \cline{7-9} \cline{11-13}  
\textbf{Description}             &               & \textbf{Implementation}                             &  & \textbf{Code} &             & \textbf{\begin{tabular}[c]{@{}r@{}}Reported\end{tabular}} &                          & \textbf{\begin{tabular}[c]{@{}r@{}}Reproduced\end{tabular}} &               & \textbf{Reported} &                                   & \textbf{\begin{tabular}[c]{@{}r@{}}Reproduced\end{tabular}} \\ \hline
FCN 32                           &               & Ours using \cite{dai2016r} & & \checkmark  &                             & 0.688             &                                   & 0.688 (0.680$\pm$0.006)                                                      &               & 0.835             &                                   & 0.835 (0.831$\pm$0.003)                                                      \\
MultiResUNet                     &               & Ours using \cite{ibtehaz2020multiresunet}& & \checkmark  &              & 0.847             &                                   & 0.847 (0.824$\pm$0.010)                                                     &               & 0.919             &                                   & 0.919 (0.902$\pm$0.007)                                                      \\
2D CNN                       &               & Cheng~\cite{cheng2017volume}& &   &                & 0.865             &                                   & 0.883 (0.871$\pm$0.008)                                                                    &               & -                 &                                   & 0.938 (0.932$\pm$0.004)                                                                    \\
3D Vanilla U-Net                 &               & Ours using \cite{cciccek20163d}& & \checkmark             &               & 0.866             &                                   & 0.866 (0.857$\pm$0.006)                                                      &               & 0.929             &                                   & 0.929 (0.924$\pm$0.003)                                                      \\
Tiramisu                         &               & Ours using \cite{jegou2017one} & & \checkmark            &               & 0.872             &                                   & 0.872 (0.857$\pm$0.017)                                                      &               & 0.932             &                                   & 0.932 (0.924$\pm$0.009)                                                      \\
2D U-Net                         &               & Casser~\cite{casser2020fast} & & \checkmark &               & 0.878             &                                   & 0.865 (0.853$\pm$0.015)                                                                    &               & 0.935             &                                   & 0.930 (0.922$\pm$0.007)                                                                    \\
3D SE U-Net                      &               & Ours & & \checkmark &               & 0.879             &                                   & 0.879 (0.874$\pm$0.007)                                                      &               & 0.936             &                                   & 0.936 (0.933$\pm$0.004)                                                      \\
3D Attention U-Net               &               & Ours & & \checkmark &               & 0.880             &                                   & 0.880 (0.876$\pm$0.003)                                                      &               & 0.936             &                                   & 0.936 (0.934$\pm$0.002)                                                      \\
nnU-Net framework              &               & Isensee~\cite{isensee2021nnu} & & \checkmark &               & 0.882             &                                   & -                                                      &               & 0.938             &                                   & -                                                      \\
MNet                             &               & Ours using \cite{fu2018joint} & & \checkmark            &               & 0.883             &                                   & 0.883 (0.874$\pm$0.007)                                                      &               & 0.938             &                                   & 0.938 (0.929$\pm$0.004)                                                      \\
2D Residual U-Net                &               & Ours & & \checkmark &               & 0.885             &                                   & 0.885 (0.880$\pm$0.004)                                                      &               & 0.939             &                                   & 0.939 (0.937$\pm$0.002)                                                      \\
3D U-Net                         &               & Ours  & & \checkmark &               & 0.885             &                                   & 0.885 (0.878$\pm$0.004)                                                      &               & 0.939             &                                   & 0.939 (0.935$\pm$0.002)                                                      \\
nnU-Net              &               & Ours using~\cite{isensee2021nnu} & & \checkmark &               & 0.888             &                                   & 0.888 (0.881$\pm$0.005)                                                      &               & 0.941             &                                   & 0.941 (0.937$\pm$0.003)                                                      \\
3D Residual U-Net                &               & Ours  & & \checkmark &               & 0.888             &                                   & 0.888 (0.883$\pm$0.002)                                                      &               & 0.941             &                                   & 0.941 (0.938$\pm$0.001)                                                      \\
2D SE U-Net                      &               & Ours & & \checkmark   &               & 0.888             &                                   & 0.888 (0.882$\pm$0.003)                                                      &               & 0.941             &                                   & 0.941 (0.937$\pm$0.002)                                                      \\
U-Net++                       &               & Ours using~\cite{zhou2018unet++} & & \checkmark &               & 0.888             &                                   & 0.888 (0.884$\pm$0.003)                                                                  &               & 0.941                 &                                   & 0.941 (0.938$\pm$0.001)                                                                    \\
3D CNN                       &               & Cheng~\cite{cheng2017volume} & &  &               & 0.889             &                                   & 0.800 (0.738$\pm$0.034)                                                                  &               & -                 &                                   & 0.894 (0.860$\pm$0.018)                                                                    \\
2D U-Net+Z-filtering             &               & Casser~\cite{casser2020fast} & & \checkmark &               & 0.890             &                                   & 0.870 (0.858$\pm$0.015)                                                                   &               & 0.942             &                                   & 0.931 (0.925$\pm$0.007)                                                                    \\
FCN 8                            &               & Ours using \cite{dai2016r} & & \checkmark            &               & 0.893             &                                   & \textbf{0.893 (0.888$\pm$0.002)}                                                      &               & \textbf{0.943}             &                                   & \textbf{0.943 (0.941$\pm$0.001)}                                                      \\
2D U-Net                         &               & Ours & & \checkmark   &               & 0.893             &                                   & \textbf{0.893 (0.888$\pm$0.002)}                                                      &               & 0.942             &                                   & 0.942 (0.941$\pm$0.001)                                                      \\
2D Attention U-Net      &      & Ours & & \checkmark   &      & 0.893    &                          & \textbf{0.893 (0.890$\pm$0.002)}                                             &      & \textbf{0.943}    &                          & \textbf{0.943 (0.942$\pm$0.001)}                                             \\
3D U-Net                         &               & Xiao~\cite{xiao2018automatic}  & &   &               & 0.900             &                                   & 0.881 (0.875$\pm$0.003)                                                                    &               & -                 &                                   & 0.937 (0.934$\pm$0.002)                                                                    \\
CNN+3 Post-proc.        &     & Oztel~\cite{oztel2017mitochondria}& &  &      & \textbf{0.907}    & \textbf{}                         & 0.683 (0.599$\pm$0.067)                                                      &      & -        &                          & 0.800 (0.757$\pm$0.106)                                                           \\
 \hline
\end{tabular}

}
        
        \captionof{table}{Reported vs. reproduced scores in the Lucchi dataset. The \textit{Reported} columns correspond to the best score claimed by authors of published works or the maximum score obtained during our experiments for entries without associated publication. The \textit{Reproduced} columns contain the maximum, mean and standard deviation values obtained while reproducing each corresponding method. Best scores of each column are presented in bold.}
\label{table:lucchi_final_table}
    \end{minipage}
\end{sideways}\par}

\subsubsection{Comparison with reported results}

To expose the sometimes large differences between the metric values reported in publications and those obtained by reproducing the very same methods, we have summarized in Table~\ref{table:lucchi_final_table} the results of the top-performing published methods, together with those of state-of-the-art approaches and our proposed networks. Notice all reproduced values correspond to the best configuration found, i.e., using the optimal pre-processing, architecture, output reconstruction and post-processing strategies for each method. The availability of original code, including that of the present paper, is also indicated in the table. Following the traditional reporting of the results, the table is ordered by increasing value of reported foreground IoU score. Notice the gap between the averaged IoU and the reported value increases with the standard deviation, underling the importance of finding stable configurations (with low standard deviations) so as not to depend on a large computation budget~\cite{dodge2019show}. 

Our proposed 2D U-Net and 2D Attention U-Net models, together with the FCN8 model reached the highest reproducible foreground IoU score with a value of $0.893$. In particular, the 2D Attention U-Net achieved an slightly higher average score in a very consistent manner. Best values were obtained using blending and ensemble for output reconstruction and Z-filtering as post-processing (see Figure \ref{fig:predictions} for an example of some of the proposed networks' predictions). As opposed to other approaches shown in Table~\ref{table:lucchi_final_table}, the standard deviation of our results is consistently low, guaranteeing good performance and reducing the number of experiments needed to reach optimal segmentations.

As expected, the lack of code associated with a publication enormously hinders the reproduction of the claimed results. Interestingly, in the case of the 2D approach from Cheng \etal~\cite{cheng2017volume}, our implementation improved over their published results, stressing the benefits of optimizing the whole segmentation workflow. Notice there are two table entries for results with nnU-Net~\cite{isensee2021nnu}: one using their entire training framework, and one plugging the best architecture found by their framework into ours.

\subsubsection{Ablation Study} 
\label{sec:ablation}

To investigate the impact and relevance of each component in our proposed networks, we performed an ablation study of our 2D U-Net architecture, as the rest of the proposed models are based on it. We compared six ablated versions with incremental changes: 1) a baseline four-level 2D U-Net model containing ReLU activations, Glorot/Xavier uniform kernel initialization~\cite{glorot2010understanding}, $16$ feature maps in the first level of the network that are doubled on each level, and no regularization or DA; 2) the baseline with basic DA (random rotations and horizontal and vertical flips); 3) adding dropout as regularization method; 4) using ELU as activation function ($\alpha=1$); 5) using \textit{He normal}~\cite{he2015delving} as kernel initialization; 6) adding attention gates~\cite{SCHLEMPER2019197} in the skip connections.

\begin{table}[]
\ra{1.2}
\begin{tabular}{rrrrrrr}
\hline
\multicolumn{1}{l}{}                                                                                                                                        & \multicolumn{1}{l}{} & \multicolumn{5}{c}{\textbf{Foreground IoU}}                                                                    \\ \cline{3-7} 
\textbf{Method}                                                                                                                                             & \phantom{abc}        & \textbf{Per Patch}       & \phantom{abc} & \textbf{50\% Overlap}    & \phantom{abc} & \textbf{Full Image}      \\ \hline
Baseline - 2D U-Net
&                      & 0.725$\pm$0.020          &               & 0.748$\pm$0.027          &               & 0.739$\pm$0.002          \\ \hline
+ DA                                                                                                                                                        &                      & 0.856$\pm$0.007          &               & 0.872$\pm$0.003          &               & 0.871$\pm$0.004          \\
+ Dropout                                                                                                                                                   &                      & 0.867$\pm$0.003          &               & 0.880$\pm$0.002          &               & 0.881$\pm$0.002          \\
+ ELU activation                                                                                                                                            &                      & 0.873$\pm$0.003          &               & 0.880$\pm$0.001          &               & 0.881$\pm$0.002          \\
+ He initializer                                                                                                                                            &                      & 0.873$\pm$0.003          &               & 0.880$\pm$0.002          &               & 0.881$\pm$0.003          \\
+ Attention Gates                                                                                                                                           &                      & \textbf{0.875$\pm$0.003} &               & \textbf{0.882$\pm$0.003} &               & \textbf{0.884$\pm$0.002} \\ \hline
\end{tabular}
\vspace{0.2cm}
\caption{Ablation study or our full 2D model. From the top to the bottom, on each row, incremental modifications are applied based on the previous configuration. \textit{DA} refers to applying data augmentation to the \textit{Baseline} configuration, \textit{Dropout} corresponds to applying dropout to the \textit{Baseline+DA} configuration and so on.}
\label{table:ablation}
\end{table}

The quantitative evaluation results on the Lucchi dataset for each case are shown in Table~\ref{table:ablation}, where we divide the results using different evaluation frameworks introduced in Section \ref{sec:eval_uncertainties}. As it can be appreciated, the IoU values vary significantly if they are provided by patch or by reconstructing the final output, highlighting once more the need of specifying the framework chosen when presenting the results. 

As can be seen, the use of DA, together with dropout, clearly outperforms the baseline architecture by a large margin. In the same way, the usage of ELU improves over the use of ReLU activation functions. Conversely, changing the kernel initialization from Glorot uniform to He normal has marginal effects in the final result, so either can be used. Finally, introducing attention in the skip connections, as suggested in~\cite{SCHLEMPER2019197}, helped increasing the network performance and maintaining results stability.

\subsection{Results on Lucchi++ and Kasthuri++}
\label{sec:lucchi++_kasthuri++}

Aiming to test how well the best solutions found for Lucchi would generalize in other datasets, we applied the same exact configurations to the Lucchi++ and Kasthuri++ datasets (see Section \ref{sec:other_sota_networks}) and compared their performance with that reported by \cite{casser2020fast}. A full summary of the results is shown in Table~\ref{table:lucchi++_kasthuri++}, where we can see our models outperform all previously reported results by a large margin.

Notice the Kasthuri++ dataset is anisotropic and contains lower resolution in the z-axis. Therefore, we modified our proposed 3D networks by removing the downsampling in that axis in their pooling operations and using shallower architectures (three levels instead of four). A full description of the configurations tested can be found in the supplementary material.

\begin{table}[]
\hspace{-1.2cm}
\begin{adjustbox}{max width=1.2\textwidth}
\begin{tabular}{lrrcrcrrrcrr}
\hline
                    & \multicolumn{1}{l}{}          & \multicolumn{1}{l}{} & \multicolumn{1}{l}{}         & \multicolumn{1}{l}{} & \multicolumn{3}{c}{\textbf{Foreground IoU}}                                           & \multicolumn{1}{l}{} & \multicolumn{3}{c}{\textbf{Overall IoU}}                                              \\ \cline{6-8} \cline{10-12} 
\textbf{Dataset}    & \textbf{Description}          & \phantom{abc}        & \textbf{Author}              & \phantom{abc}        & \textbf{Maximum} & \phantom{abc}        & \multicolumn{1}{c}{\textbf{(mean$\pm$std)}} & \phantom{abc}        & \textbf{Maximum} & \phantom{abc}        & \multicolumn{1}{c}{\textbf{(mean$\pm$std)}} \\ \hline
\textbf{Lucchi++}   & 2D U-Net                      &                      & Casser~\cite{casser2020fast} &                      & 0.888            &                      & -                                           &                      & 0.940            &                      & -                                           \\
                    & 2D U-Net+Z Filtering          &                      & Casser~\cite{casser2020fast} &                      & 0.900            &                      & -                                           &                      & 0.946            &                      & -                                           \\
                    & 2D Residual U-Net ($\dag$)    &                      & Ours                         &                      & 0.908            &                      & 0.904$\pm$0.004                             &                      & 0.943            &                      & 0.948$\pm$0.002                             \\
                    & 2D U-Net ($\dag$)             &                      & Ours                         &                      & 0.916            &                      & 0.911$\pm$0.006                             &                      & 0.955            &                      & 0.952$\pm$0.003                             \\
                    & 2D Attention U-Net ($\dag$)   &                      & Ours                         &                      & 0.919            &                      & 0.914$\pm$0.003                             &                      & 0.956            &                      & 0.954$\pm$0.001                             \\
                    & 3D U-Net (\Admetos)           &                      & Ours                         &                      & 0.923            &                      & 0.915$\pm$0.007                             &                      & 0.958            &                      & 0.954$\pm$0.004                             \\
                    & 3D Attention U-Net (\Admetos) &                      & Ours                         &                      & 0.923            &                      & 0.912$\pm$0.008                             &                      & 0.959            &                      & 0.953$\pm$0.004                             \\
                    & 3D Residual U-Net (\Admetos)  &                      & Ours                         &                      & \textbf{0.926}   &                      & 0.919$\pm$0.005                             &                      & \textbf{0.960}   &                      & 0.957$\pm$0.003                             \\ \hline
\textbf{Kasthuri++} & 2D U-Net                      &                      & Casser~\cite{casser2020fast} &                      & 0.845            &                      & -                                           &                      & 0.920            &                      & -                                           \\
                    & 2D U-Net+Z Fil.               &                      & Casser~\cite{casser2020fast} &                      & 0.846            &                      & -                                           &                      & 0.920            &                      & -                                           \\
                    & 2D Residual U-Net (\Admetos)  &                      & Ours                         &                      & 0.908            &                      & 0.906$\pm$0.001                             &                      & 0.953            &                      & 0.950$\pm$0.001                             \\
                    & 2D Attention U-Net (\Admetos) &                      & Ours                         &                      & 0.915            &                      & 0.913$\pm$0.001                             &                      & 0.956            &                      & 0.954$\pm$0.001                             \\
                    & 2D U-Net (\Admetos)           &                      & Ours                         &                      & 0.916            &                      & 0.913$\pm$0.002                             &                      & 0.955            &                      & 0.954$\pm$0.001                             \\
                    & 3D U-Net (\Admetos)           &                      & Ours                         &                      & 0.934            &                      & 0.932$\pm$0.001                             &                      & 0.965            &                      & 0.965$\pm$0.001                             \\
                    & 3D Residual U-Net (\Admetos)  &                      & Ours                         &                      & 0.934            &                      & 0.933$\pm$0.001                             &                      & 0.966            &                      & 0.966$\pm$0.000                             \\
                    & 3D Attention U-Net (\Admetos) &                      & Ours                         &                      & \textbf{0.937}   &                      & 0.934$\pm$0.001                             &                      & \textbf{0.967}   &                      & 0.966$\pm$0.001                             \\ \hline
                    & \multicolumn{1}{l}{}          & \multicolumn{1}{l}{} & \multicolumn{1}{l}{}         & \multicolumn{1}{l}{} &                  & \multicolumn{1}{l}{} & \multicolumn{1}{l}{}                        & \multicolumn{1}{l}{} &                  & \multicolumn{1}{l}{} & \multicolumn{1}{l}{}                        \\
\multicolumn{12}{c}{($\dag$) 0\% overlap output reconstruction, blended ensemble and z-filtering post-processing}                                                                                                                                                                                                                              \\
\multicolumn{12}{c}{(\Admetos) 50\% overlap output reconstruction and ensemble post-processing}                                                                                                                                                                                                        
\end{tabular}
\end{adjustbox}

\vspace{0.2cm}
\caption{Results obtained in the Lucchi++ and Kasthuri++ datasets.}
\label{table:lucchi++_kasthuri++}
\end{table}

\section{Conclusions}
\label{conclusions}

By a complete experimental study of state-of-the-art deep learning models with modern training workflows, we have revealed significant problems of reproducibility in the domain of mitochondria segmentation in EM data. Moreover, by disentangling the effects of novel architectures from those of the training choices (i.e., pre-processing, data augmentation, output reconstruction and post-processing strategies) over a set of multiple executions of the same configurations, we have found stable lightweight models that consistently lead to state-of-the-art results on the existing public datasets.

Have novel methods reached human performance? To answer that question, Casser \etal~\cite{casser2020fast} compared the results of human annotators in the Lucchi dataset, producing a foreground IoU value of $0.884$. This would suggest that many of the models presented in Table \ref{table:lucchi_final_table} outperform indeed humans in this task. Nevertheless, all methods fell short of the threshold of $0.91$ of foreground IoU, what could be due to the annotation inconsistencies discussed in Section \ref{sec:other_sota_networks}. To investigate further, we created two slightly different versions of the mitochondria ground truth labels by 1-pixel morphological dilation and erosion. The foreground IoU value of the resulting labels against the original ground truth was $0.885$ for the dilated version and $0.904$ for the eroded one. Thus, this enforces the idea that the dataset is not pixel-level accurate, so it could be argued that all the methods with IoU values within a range of $0.009$ or less can probably be considered to have similar performance. The same experiment was done with the ground truth labels of Lucchi++ and Kasthuri++. The foreground IoU values obtained dilating and eroding Lucchi++ were $0.898$ and $0.919$ respectively, while $0.927$ and $0.922$ were obtained in Kasthuri++. Indeed, even the average score of many of our models outperform those values (see Table \ref{table:lucchi++_kasthuri++}). This suggests the performance on all three datasets has probably saturated, as new architectures and training frameworks cannot improve beyond the limits inherent to semantic segmentation and the size of the datasets.

In closing, we believe further progress in mitochondria segmentation in EM will require (1) larger and more complex datasets~\cite{wei2020mitoem}, and (2) the adoption of a reproducibility checklist or set of best practices~\cite{dodge2019show} to report more comprehensive results and allow robust future comparisons.

\section{Acknowledgements}
\label{acknowledgements}

We acknowledge the support of Ministerio de Ciencia, Innovación y Universidades, Agencia Estatal de Investigación, under Grants TEC2016-78052-R and PID2019-109820RB-I00, MINECO/FEDER, UE, co-financed by European Regional Development Fund (ERDF), “A way of making Europe”. Work produced with the support of a 2020 Leonardo Grant for Researchers and Cultural Creators, BBVA Foundation.
\bibliographystyle{splncs04}
\bibliography{document}

\appendix

\renewcommand\thetable{\thesection.\arabic{table}}    
\setcounter{table}{0}

\renewcommand\thefigure{\thesection.\arabic{figure}}    
\setcounter{figure}{0}

\newpage
\section{Appendix: Network predictions on Lucchi}

\begin{figure}[H]
\vspace{-0.7cm}
\centering
\begin{subfigure}{0.9\textwidth}
  \includegraphics[width=\linewidth]{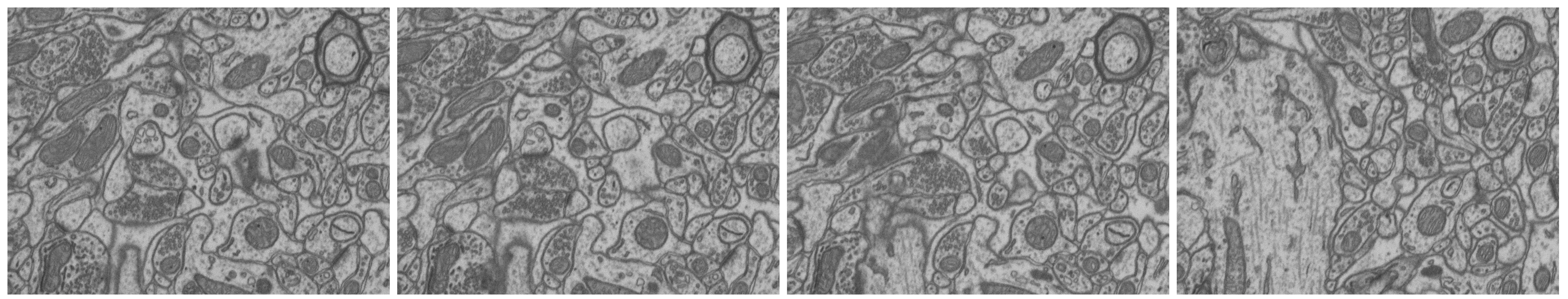}
  \vspace{-0.55cm}
  \caption{Original images}
\end{subfigure}
\begin{subfigure}{0.9\textwidth}
  \includegraphics[width=\linewidth]{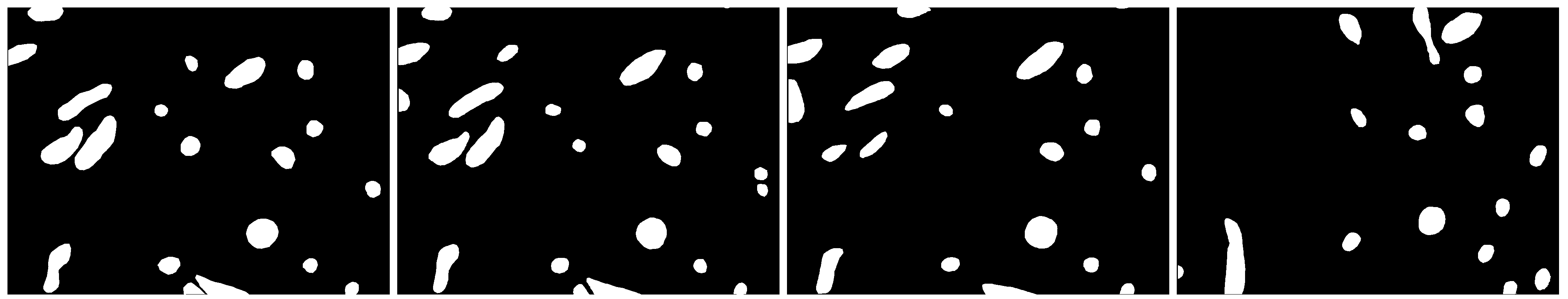}
  \vspace{-0.55cm} 
  \caption{Ground Truth}
\end{subfigure}
\begin{subfigure}{0.9\textwidth}
  \includegraphics[width=\linewidth]{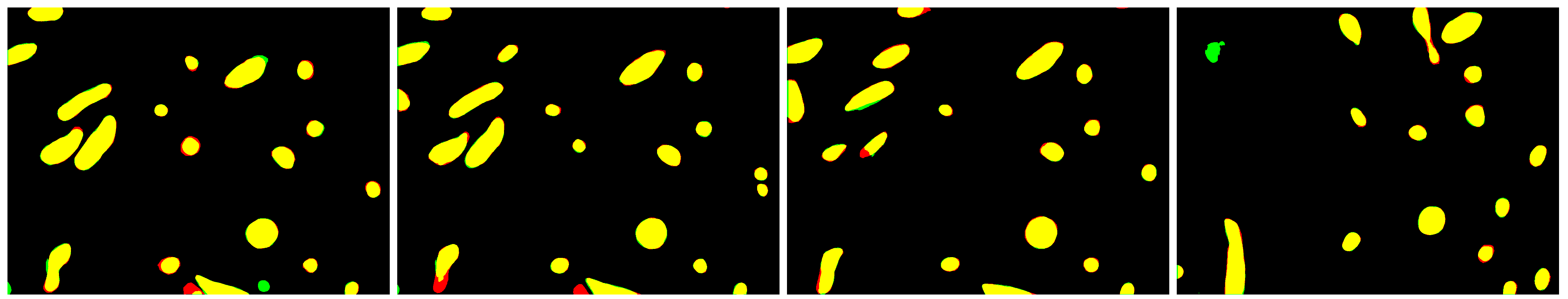}
  \vspace{-0.55cm}
  \caption{Basic 2D U-Net predictions}
\end{subfigure}
\begin{subfigure}{0.9\textwidth}
  \includegraphics[width=\linewidth]{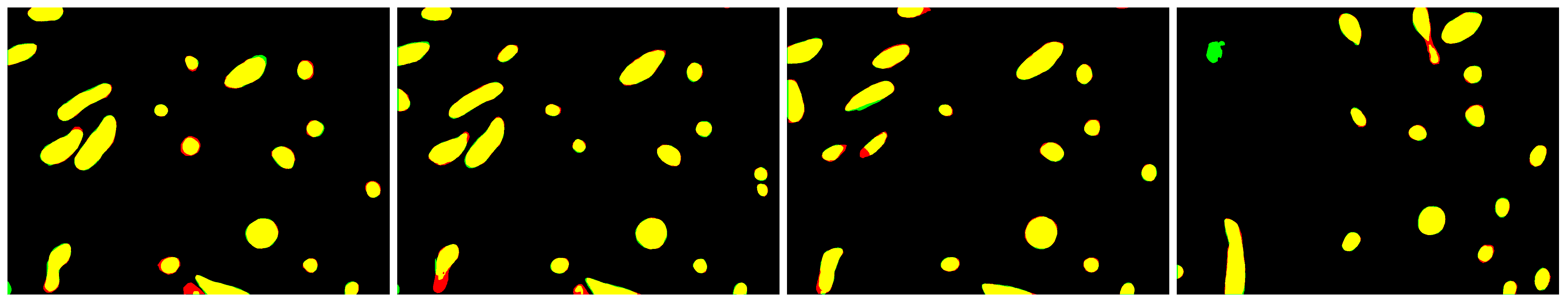}
  \vspace{-0.55cm}
  \caption{2D Attention U-Net predictions}
\end{subfigure}
\begin{subfigure}{0.9\textwidth}
  \includegraphics[width=\linewidth]{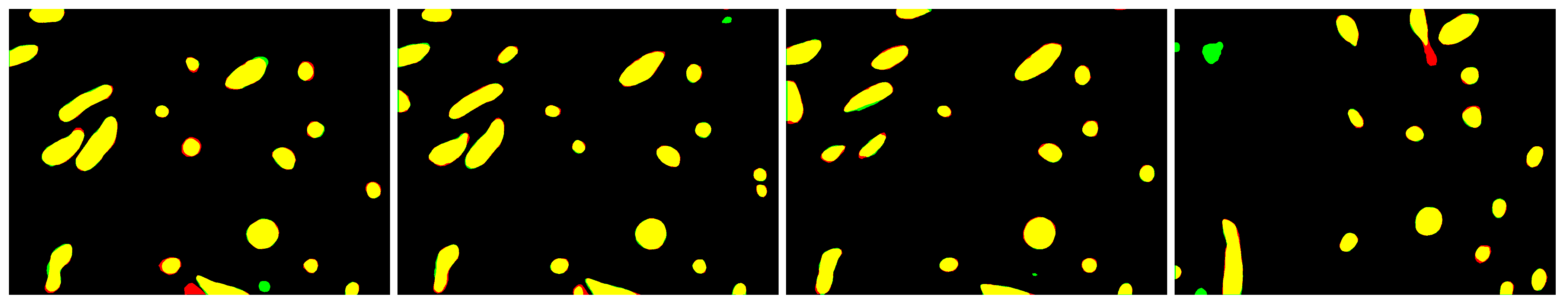}
  \vspace{-0.55cm}
  \caption{2D Residual U-Net predictions}
\end{subfigure}
\begin{subfigure}{0.9\textwidth}
  \includegraphics[width=\linewidth]{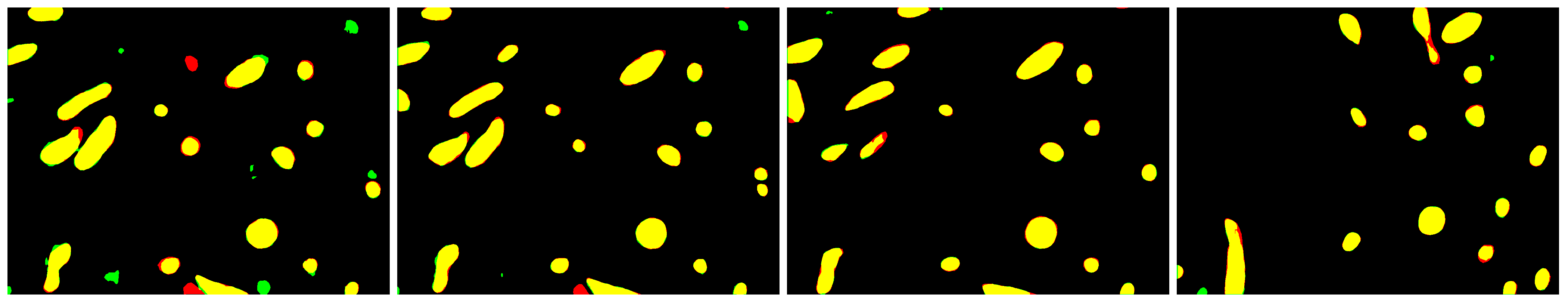}
  \vspace{-0.55cm}
  \caption{Basic 3D U-Net predictions}
\end{subfigure}
\begin{subfigure}{0.9\textwidth}
  \includegraphics[width=\linewidth]{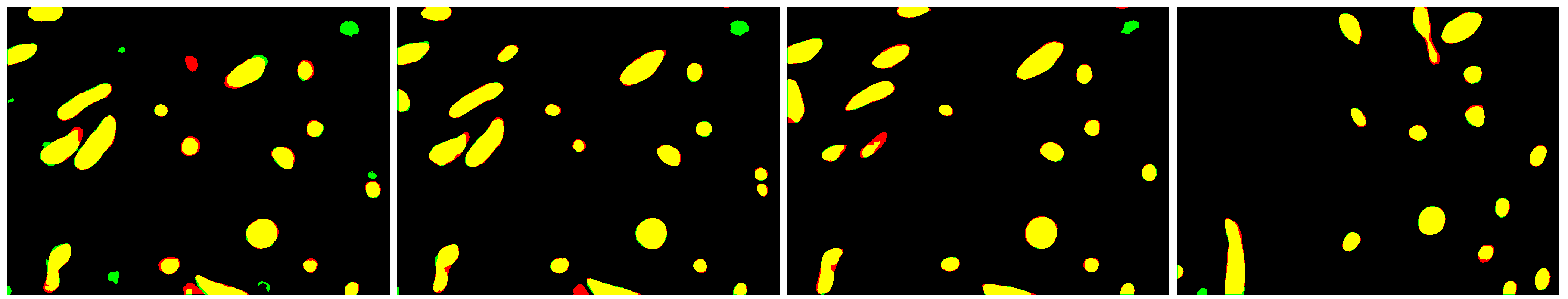}
  \vspace{-0.55cm}
  \caption{3D Attention U-Net predictions}
\end{subfigure}
\begin{subfigure}{0.9\textwidth}
  \includegraphics[width=\linewidth]{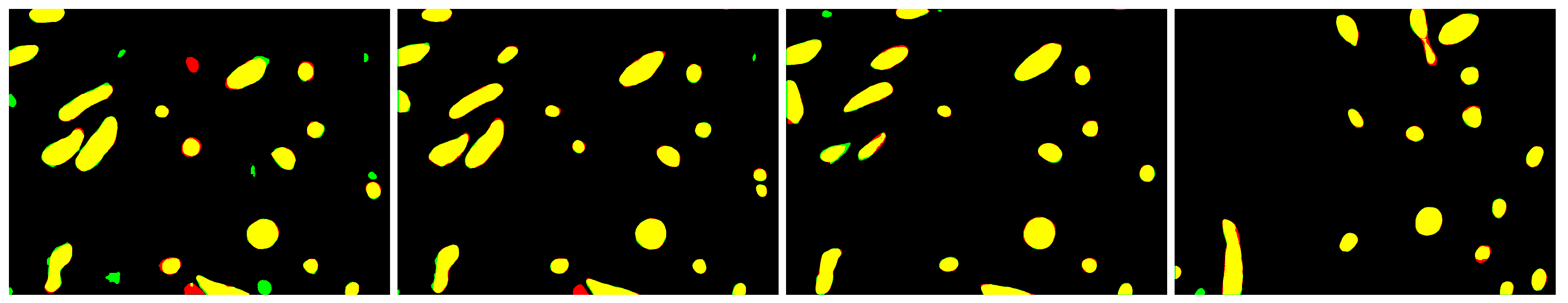}
  \vspace{-0.55cm}
  \caption{3D Residual U-Net predictions}
\end{subfigure}
\vspace{-0.2cm}
\caption{Predictions of the configuration that measures the best IoU score for each proposed network on Lucchi dataset. Yellow: True Positive; Green: False Positive; Red: False Negative.}
\label{fig:predictions}
\end{figure}

\section{Appendix: Hyperparameter search space for each configuration}

\subsection{Notations}
\begin{itemize}
    \item $[a, b]$ : Range between two possible values. E.g. \textit{zoom([0.75,1.25])} correspond to random zoom value between $0.75$ and $1.25$.
    \item $[a, b, c]$ : All values from $a$ to $b$ with $c$ step. E.g. $[10, 300, 10]$ correspond to $10, 20, 30, 40, ..., 300$. 
    \item $choice[a, b, ...]$ : one value between $a$, $b$ and so on. E.g. \textit{[10, 15, 20, 30, 60]} possible values are: $10$ or $15$ or $20$ or $30$ or $60$ (but only one). 
    \item $a, b, c, ...$ : all tested values. E.g. flips, rotations, etc.
\end{itemize}

\subsection{Training setup}
\begin{table}[]
\begin{tabular}{lll}
Computing Infrastructure & \phantom{abc} & GeForce GTX 1080 Ti                   \\
GPU libraries & \phantom{abc} & CUDA (10.1) + cuDNN (7.6.5)                   \\
Operating System & \phantom{abc} & Ubuntu 16.04.6 LTS                           \\
Number of runs           & & 10                                            \\
Implementation           & & Tensorflow (2.1.0) + Keras (2.2.4-tf)                            \\
Code     & & \href{https://github.com/danifranco/EM_Image_Segmentation}{Github}  \\
Math seed used           & & 42 \\
Environment used & & Anaconda $\rightarrow$ \href{https://github.com/danifranco/EM_Image_Segmentation/blob/master/env/DL_EM_base_env.yml}{DL\_EM\_base\_env}
\end{tabular}
\end{table}

\newpage

\subsection{2D U-Net}

\begin{table}[]
\vspace{-0.6cm}
%
\end{adjustbox}
\vspace{0.2cm}
\caption{Hyperparameter search space for the proposed 2D U-Net.}
\end{table}

\newpage

\subsection{2D Residual U-Net}

\begin{table}[]
\vspace{-0.6cm}
\begin{tabular}{ll}
Template to reproduce the results: & \href{https://github.com/danifranco/EM_Image_Segmentation/blob/master/templates/Residual_U-Net_2D_template.py}{Residual\_U-Net\_2D\_template.py}                   \\
\end{tabular}%
\vspace{0.2cm}
\caption{Hyperparameter search space for the proposed 2D Residual U-Net.}
\end{table}

\newpage

\subsection{3D U-Net}

\begin{table}[]
\vspace{-0.6cm}
%
\end{adjustbox}
\vspace{0.2cm}
\caption{Hyperparameter search space for proposed 3D U-Net.}
\end{table}

\newpage
\subsection{3D Residual U-Net}

\begin{table}[]
%
\end{adjustbox}

\vspace{0.2cm}
\caption{Hyperparameter search space for proposed 3D Residual U-Net.}
\end{table}

\newpage
\subsection{Cheng 2D~\cite{cheng2017volume}}

\begin{table}[]
%
                 \\
Patch shape                 &               & $256\times256$                      &               & $256\times256$                      \\
Probability map            &               & \textit{choice}[False,True]                 &               & True                                \\
Probability for each class &               & Foreground: 0.94 ; Background: 0.06 &               & Foreground: 0.94 ; Background: 0.06 \\ \hline
Data augmentation          &               & flips, rotation\_range([-180,180])       &               & flips, rotation\_range([-180,180])       \\ \hline
Number of epochs           &               & \textit{choice}[4000,400]                   &               & 400                                 \\
Patience                   &               & 200                                 &               & 200                                 \\
Batch size                 &               & 24                                  &               & 24                                  \\
Loss type                  &               & BCE                                 &               & BCE                                 \\
Optimizer                  &               & \textit{choice}[SGD,Adam]                          &               & Adam                                 \\
SGD learning rate          &               & \textit{choice}[0.002,0.05]                  &               & -                                   \\
Adam learning rate         &               & \textit{choice}[0.0001]                      &               & 0.0001                              \\
learning rate scheduler    &               & \textit{choice}[True,False]                  &               & False                               \\
Dropout                    &               & \textit{choice}[0,0.1]                      &               & 0                                   \\ \hline
\end{tabular}
\vspace{0.2cm}
\caption{Hyperparameter search space for 2D network proposed by Cheng \textit{et al.}~\cite{cheng2017volume}.}
\end{table}

\newpage
\subsection{Cheng 3D~\cite{cheng2017volume}}

\begin{table}[]
\begin{tabular}{l}
Templates to reproduce the results: \\
\textbf{~~~~\textit{Original}:} \href{https://github.com/danifranco/EM_Image_Segmentation/blob/master/sota_implementations/cheng_2017/cheng_3D_template_V0.py}{cheng\_3D\_template\_V0.py}                   \\
\textbf{~~~~\textit{Modified}:} \href{https://github.com/danifranco/EM_Image_Segmentation/blob/master/sota_implementations/cheng_2017/cheng_3D_template_V1.py}{cheng\_3D\_template\_V1.py}                    
\end{tabular}
\end{table}

\begin{table}[]
\hspace{-2.8cm}
\begin{tabular}{rcrcr}
                        & \phantom{abc} &                                                                                                                                                  & \phantom{abc} &                                                                                        \\ \hline
\textbf{Hyperparameter} &               & \textbf{Search space}                                                                                                                            &               & \textbf{Best assignment}                                                               \\ \hline
Duplicate train         &               & \textit{choice}[1,12]                                                                                                                                    &               & 12                                                                                     \\
Validation              &               & False                                                                                                                                            &               & False                                                                                  \\ \hline
Subvolumes              &               & \begin{tabular}[c]{@{}r@{}}\textit{choice}[created from train data before\\network training, random selection from\\the whole data during training]\end{tabular} &               & \begin{tabular}[c]{@{}r@{}}created from training data\\ before network training\end{tabular} \\
Subvolume shape         &               & $128\times128\times96$                                                                                                                           &               & $128\times128\times96$                                                                 \\
Probability map         &               & \textit{choice}[False,True]                                                                                                                              &               & False                                                                                  \\ \hline
Data augmentation       &               & \begin{tabular}[c]{@{}r@{}}flips,\\square\_rotations([0,90,180,270]),\\elastic  \end{tabular}                                                                                                                      &               & square\_rotations([0,90,180,270])                                                             \\ \hline
Number of epochs        &               & \textit{choice}[545,150]                                                                                                                                 &               & 150                                                                                    \\
Patience                &               & \textit{choice}[50,200]                                                                                                                                  &               & 50                                                                                     \\
Batch size              &               & \textit{choice}[1,3]                                                                                                                                      &               & 3                                                                                      \\
Loss type               &               & BCE                                                                                                                                              &               & BCE                                                                                    \\
Optimizer               &               & \textit{choice}[SGD,Adam]                                                                                                                                        &               & Adam                                                                                   \\
SGD learning rate       &               & 0.1                                                                                                                                              &               & -                                                                                      \\
Adam learning rate      &               & 0.0001                                                                                                                                           &               & 0.0001                                                                                 \\
learning rate scheduler &               & \textit{choice}[True,False]                                                                                                                               &               & False                                                                                  \\
Dropout                 &               & 0.1                                                                                                                                              &               & 0.1                                                                                    \\ \hline
\end{tabular}
\vspace{0.2cm}
\caption{Hyperparameter search space for 3D network proposed by Cheng \textit{et al}.~\cite{cheng2017volume}.}
\end{table}

\newpage

\subsection{Casser~\cite{casser2020fast}}

\begin{table}[]
\begin{tabular}{l}
Templates to reproduce the results: \\
\textbf{~~~~\textit{Original}:} \href{https://github.com/danifranco/EM_Image_Segmentation/blob/master/sota_implementations/casser_2018/casser_template_V0.py}{casser\_template\_V0.py}                   \\
\textbf{~~~~\textit{Modified}:} \href{https://github.com/danifranco/EM_Image_Segmentation/blob/master/sota_implementations/casser_2018/casser_template_V1.py}{casser\_template\_V1.py}                    
\end{tabular}
\end{table}

\begin{table}[]
\vspace{-1.8cm}
\hspace{-2.5cm}
\begin{tabular}{rcrcr}
                           & \phantom{abc} &                                                                                                                                                                                                             & \phantom{abc} &                                                                                              \\ \hline
\textbf{Hyperparameter}    &               & \textbf{Search space}                                                                                                                                                                                       &               & \textbf{Best assignment}                                                                     \\ \hline
Duplicate train            &               & \textit{choice}[1,2,12]                                                                                                                                                                                            &               & 2                                                                                            \\
Validation                 &               & True                                                                                                                                                                                                        &               & True                                                                                         \\
Random validation          &               & \textit{choice}[True,False]                                                                                                                                                                                         &               & False                                                                                        \\
\% of train as validation  &               & \textit{choice}[5\%,10\%,20\%,30\%]                                                                                                                                                                                  &               & 10\%                                                                                         \\ \hline
Patches                    &               & \begin{tabular}[c]{@{}r@{}}random selection from the \\ whole data during train\end{tabular}                                                                                                                &               & \begin{tabular}[c]{@{}r@{}}random selection from the \\ whole data during train\end{tabular} \\
Patch size                 &               & $512\times512$                                                                                                                                                                                              &               & $512\times512$                                                                               \\
Probability map            &               & \textit{choice}[False,True]                                                                                                                                                                                         &               & True                                                                                         \\
Probability for each class &               & \begin{tabular}[c]{@{}r@{}}Foreground: 0.9 ; Background: 0.1,\\ Foreground: 0.94 ; Background: 0.06,\end{tabular}                                                                                           &               & Foreground: 0.94 ; Background: 0.06                                                          \\ \hline
Data augmentation          &               & \begin{tabular}[c]{@{}r@{}}flips,\\ square\_rotations([0,90,180,270]),\\ rotation\_range([0,180]),\\
shift([0.1,0.3]),\\
shearing([0.1,0.3]),\\
brightness\_range([0.8,1.2]), \\ median\_filtering(size=5)\end{tabular} &               & \begin{tabular}[c]{@{}r@{}}flips,\\rotation\_range([0,180])\end{tabular}                                                                \\ \hline
Number of epochs           &               & 360                                                                                                                                                                                                         &               & 360                                                                                          \\
Patience                   &               & \textit{choice}[50,200]                                                                                                                                                                                             &               & 200                                                                                          \\
Batch size                 &               & \textit{choice}[4,6]                                                                                                                                                                                                &               & 4                                                                                            \\
Loss type                  &               & BCE                                                                                                                                                                                                         &               & BCE                                                                                          \\
Optimizer                  &               & \textit{choice}[SGD, Adam]                                                                                                                                                                                         &               & Adam                                                                                         \\
SGD learning rate          &               & \textit{choice}[0.001,0.002,0.005,0.008,0.01]                                                                                                                                                                                               &               & -                                                                                            \\
Adam learning rate         &               & \textit{choice}[0.0005,0.0001,0.001]                                                                                                                                                                                              &               & 0.0005                                                                                       \\
Dropout                    &               & 0.2                                                                                                                                                                                                         &               & 0.2                                                                                          \\ \hline
\end{tabular}
\vspace{0.2cm}
\caption{Hyperparameter search space for network proposed by Casser \textit{et al.}~\cite{casser2020fast}.}
\end{table}

\newpage
\subsection{Xiao~\cite{xiao2018automatic}}

\begin{table}[]
\begin{tabular}{l}
Templates to reproduce the results: \\
\textbf{~~~~\textit{Original}:} \href{https://github.com/danifranco/EM_Image_Segmentation/blob/master/sota_implementations/xiao_2018/xiao_template_V0.py}{xiao\_template\_V0.py}                   \\
\textbf{~~~~\textit{Modified}:} \href{https://github.com/danifranco/EM_Image_Segmentation/blob/master/sota_implementations/xiao_2018/xiao_template_V1.py}{xiao\_template\_V1.py}           
\end{tabular}%
\vspace{0.2cm}
\caption{Hyperparameter search space for network proposed by Xiao \textit{et al.}~\cite{xiao2018automatic}.}
\end{table}

\newpage
\subsection{Oztel~\cite{oztel2017mitochondria}}

\begin{table}[]
\begin{tabular}{l}
Templates to reproduce the results: \\
\textbf{~~~~\textit{Original}:} \href{https://github.com/danifranco/EM_Image_Segmentation/blob/master/sota_implementations/oztel_2017/oztel_template_V0.py}{oztel\_template\_V0.py}                   \\
\textbf{~~~~\textit{Modified}:} \href{https://github.com/danifranco/EM_Image_Segmentation/blob/master/sota_implementations/oztel_2017/oztel_template_V0.py}{oztel\_template\_V1.py}                   \\
\end{tabular}%
\vspace{0.2cm}
\caption{Hyperparameter search space for network proposed by Oztel \textit{et al.}~\cite{oztel2017mitochondria}.}
\end{table}

\newpage
\subsection{FCN~\cite{long2015fully}}

\begin{table}[]
\begin{tabular}{l}
Templates to reproduce the results: \\
\textbf{~~~~FCN32:} \href{https://github.com/danifranco/EM_Image_Segmentation/blob/master/templates/FCN32_template.py}{FCN32\_template.py}                   \\
\textbf{~~~~FCN8:} \href{https://github.com/danifranco/EM_Image_Segmentation/blob/master/templates/FCN8_template.py}{FCN8\_template.py}           
\end{tabular}%
\vspace{0.2cm}
\caption{Hyperparameter search space for FCN32 anf FCN8 networks~\cite{long2015fully}.}
\end{table}

\newpage
\subsection{Tiramisu~\cite{jegou2017one}}

\begin{table}[]
\begin{tabular}{ll}
Template to reproduce the results: & \href{https://github.com/danifranco/EM_Image_Segmentation/blob/master/templates/Tiramisu_template.py}{Tiramisu\_template.py}                   \\
\end{tabular}%
\vspace{0.2cm}
\caption{Hyperparameter search space for Tiramisu~\cite{jegou2017one}.}
\end{table}

\newpage
\subsection{2D SE U-Net 2D}

\begin{table}[]
\begin{tabular}{ll}
Template to reproduce the results: & \href{https://github.com/danifranco/EM_Image_Segmentation/blob/master/templates/SE_U-Net_2D_template.py}{SE\_U-Net\_2D\_template.py}                   \\
\end{tabular}%
\vspace{0.2cm}
\caption{Hyperparameter search space for 2D U-Net (adding SE blocks~\cite{hu2018squeeze}).}
\end{table}

\newpage
\subsection{3D SE U-Net}

\begin{table}[]
\begin{tabular}{ll}
Template to reproduce the results: & \href{https://github.com/danifranco/EM_Image_Segmentation/blob/master/templates/SE_U-Net_3D_template.py}{SE\_U-Net\_3D\_template.py}                   \\
\end{tabular}%
\vspace{0.2cm}
\caption{Hyperparameter search space for 3D U-Net (adding SE blocks~\cite{hu2018squeeze}).}
\end{table}

\newpage
\subsection{MultiResUNet~\cite{ibtehaz2020multiresunet}}

\begin{table}[]
\begin{tabular}{ll}
Template to reproduce the results: & \href{https://github.com/danifranco/EM_Image_Segmentation/blob/master/templates/MultiResUNet_template.py}{MultiResUNet\_template.py}                   \\
\end{tabular}%
\vspace{0.2cm}
\caption{Hyperparameter search space for MultiResUNet~\cite{ibtehaz2020multiresunet}.}
\end{table}

\newpage
\subsection{MNet~\cite{fu2018joint}}

\begin{table}[]
\begin{tabular}{ll}
Template to reproduce the results: & \href{https://github.com/danifranco/EM_Image_Segmentation/blob/master/templates/MNet_template.py}{MNet\_template.py}                   \\
\end{tabular}%
\vspace{0.2cm}
\caption{Hyperparameter search space for MNet~\cite{fu2018joint}.}
\end{table}

\newpage
\subsection{3D Vanilla U-Net~\cite{cciccek20163d}}

\begin{table}[]
\begin{tabular}{ll}
Template to reproduce the results: & \href{https://github.com/danifranco/EM_Image_Segmentation/blob/master/templates/Vanilla_U-Net_3D_template.py}{Vanilla\_U-Net\_3D\_template.py}                   \\
\end{tabular}%
\vspace{0.2cm}
\caption{Hyperparameter search space for 3D Vanilla U-Net~\cite{cciccek20163d}.}
\end{table}

\newpage
\subsection{nnU-Net~\cite{isensee2021nnu}}

\begin{table}[]
\begin{tabular}{ll}
Template to reproduce the results:  & \href{https://github.com/danifranco/EM_Image_Segmentation/blob/master/templates/nnU-Net_template.py}{nnU-Net\_template.py}                   \\
\end{tabular}%
\vspace{0.2cm}
\caption{Hyperparameter search space for nnU-Net~\cite{isensee2021nnu}.}
\end{table}

\newpage
\subsection{U-Net++~\cite{zhou2018unet++}}

\begin{table}[]
%
             &               & -                                                                                         \\ \hline
\end{tabular}
\vspace{0.2cm}
\caption{Hyperparameter search space for U-Net++~\cite{zhou2018unet++}}
\end{table}

\end{document}